\documentclass[useAMS,usenatbib]{mn2e}
\usepackage{graphicx,amsmath,multirow,amssymb}
\usepackage{natbib}
\newcommand{\comment}[1]{}

\def\simgt{\lower.5ex\hbox{$\; \buildrel > \over \sim \;$}}
\def\simlt{\lower.5ex\hbox{$\; \buildrel < \over \sim \;$}}

\title[AGB and SAGB yields of various metallicity]
{Yields of AGB and SAGB models with chemistry of low-- and high--metallicity
Globular Clusters}
\author[P.Ventura, M. Di Criscienzo, R.Carini and F.D'Antona]{P. Ventura$^{1}$\thanks{E-mail:
paolo.ventura@oa-roma.inaf.it (AVR)}, M.Di Criscienzo$^{3}$, R.Carini$^{1,2}$ and F. D'Antona$^{1}$\\
$^{1}$INAF-Osservatorio Astronomico di Roma, Via Frascati 33, Monte Porzio Catone 00040, Italy\\
$^{2}$Dipartimento di Fisica, Universit\`a di Roma ``La Sapienza'', Italy\\
$^{3}$INAF-Osservatorio Astronomico di Capodimonte, Salita Moiariello 16, Napoli 80131, Italy}
\begin{document}

\date{Accepted, Received; in original form }

\pagerange{\pageref{firstpage}--\pageref{lastpage}} \pubyear{2002}

\maketitle

\label{firstpage}

\begin{abstract}
We present yields from stars of mass in the range $M_{\odot} \leq M \leq 8M_{\odot}$
of metallicities $Z=3\times 10^{-4}$ and $Z=8\times 10^{-3}$, thus encompassing the chemistry
of low-- and high--Z Globular Clusters. The yields are based on full evolutionary 
computations, following the evolution of the stars from the pre-Main Sequence 
through the Asymptotic Giant Branch phase, until the external envelope is lost.

Independently of metallicity, stars with $M<3M_{\odot}$ are dominated by Third Dredge--Up,
thus ejecting into their surroundings gas enriched in carbon and nitrogen. Conversely, 
Hot Bottom Burning is the main responsible for the modification of the surface chemistry 
of more massive stars, whose mass exceeds $3M_{\odot}$: their gas shows traces of 
proton--capture nucleosynthesis.

The extent of Hot Bottom Burning turns out to be strongly dependent on metallicity.
Models with $Z=8\times 10^{-3}$ achieve a modest depletion of oxygen, barely reaching
$-0.3$ dex, and do not activate the Mg--Al chain. Low--Z models with $Z=3\times 10^{-4}$ 
achieve a strong nucleosynthesis at the bottom of the envelope, with a strong destruction 
of the surface oxygen and magnesium; the most extreme chemistry is reached for models
of mass $\sim 6M_{\odot}$, where $\delta$[O/Fe]$ \sim -1.2$ and $\delta$[Mg/Fe]$\sim -0.6$.
Sodium is found to be produced in modest quantities at these low Z's, because the initial
increase due to the combined effect of the second dredge--up and of $^{22}$Ne burning
is compensated by the later destruction via proton capture. A great increase by a factor
$\sim 10$ in the aluminium content of the envelope is also expected. These results
can be used to understand the role played by intermediate mass stars in the 
self--enrichment scenario of globular clusters: the results from spectroscopic investigations
of stars belonging to the second generation of clusters with different metallicity will be 
used as an indirect test of the reliability of the present yields.

The treatment of mass loss and convection are confirmed as the main uncertainties affecting
the results obtained in the context of the modeling of the thermal pulses phase. An 
indirect proof of this comes from the comparison with other investigations in the literature,
based on a different prescription for the efficiency of convection in transporting energy 
and using a different recipe to determine the mass loss rate.
\end{abstract}


\section{Introduction}
Stars with mass below $6M_{\odot}$, shortly after the end of core helium burning,
develop a degenerate core of Carbon and Oxygen, and evolve supported by two nuclear
regions, where CNO and 3$\alpha$ burning occur. Due to the narrow dimensions of the
He--burning layer, $3\alpha$ burning is not thermally stable \citep{schw65, schw67}, 
and occurs periodically, in violent episodes known as thermal pulses (hereinafter TP); 
for most of the time CNO burning is the only energy channel active in the star 
\citep{iben75}. In the Hertzprung--Russell diagram, the evolutionary tracks, after the 
excursion to the blue during the core He--burning phase, turn again to the red; this 
evolutionary phase is commonly known as Asymptotic Giant Branch (AGB).

More massive objects, with mass $6M_{\odot} < M <8M_{\odot}$\footnote{These limits in mass
partly depend on the assumption concerning the extent of the extra--mixing region from the
external border of the convective core during the hydrogen burning phase. In the present
investigation we consider a modest overshoot; if this was neglected, the range of masses
involved in the SAGB evolution would be $8M_{\odot} < M <10M_{\odot}$}, undergo a similar 
evolution, with the difference that their internal temperatures are sufficiently high to 
trigger carbon ignition in a partially degenerate region, near the stellar center 
\citep{ritossa96, ritossa99}. In these stars the inwards propagation of a convective flame 
favors the formation of a core made up of oxygen and neon. Like their lower mass 
counterparts, they also undergo a series of thermal pulses, and evolve in the Super 
Asymptotic Giant Branch (SAGB) phase \citep{siess06, siess07, siess10}.

The fate of AGB stars is to eventually loose their envelope, and evolve to the White Dwarf 
stage. The final stages of the evolutionary history of SAGBs is more uncertain. If the mass
loss rate is low, their core grows in mass until exceeding the Chandrasekhar limit, thus
favoring the conditions for core--collapse supernova, via electron capture. Alternatively,
they evolve as ONe white dwarfs \citep{poelarends08}.

The AGB evolution is rather short compared to the previous phases of core burning, but it 
proves extremely important because it is during this phase that most of the mass is lost
from the star. Understanding the evolution of the surface chemistry of these stars is
crucial for a number of topics, such as the role played by AGBs and SAGBs as dust 
producers \citep{gail99, fg06}, that was shown to depend critically on the abundances of 
the various chemical species in the surface layers of the star \citep{ventura12a, ventura12b}.

AGB and the SAGB stars have been invoked to explain the observations of stars in
Globular Clusters (GC) \citep{ventura01}. The spectroscopic and photometric results gathered 
in the last decades indicate the presence of multiple populations 
\citep{carretta09, gratton01, dantona2005, piotto07}, and that (at least) a new generation 
of stars formed from the ashes of rapidly evolving stars belonging to the original 
population of the cluster. Massive AGBs and SAGBs appear to be the most appealing 
candidates, in spite of an ongoing debate, due to the various uncertainties 
affecting the robustness of the results of AGB modeling. This is the reason why some research 
groups argued against the possibility that AGB winds could ever reproduce the chemical 
patterns observed \citep{fenner04}, whereas other investigations showed that on the 
qualitative side the most massive AGBs produce ejecta whose chemistry is in agreement 
with the anticorrelations observed \citep{vd09}.

The investigations by \citet{annibale, dercole10, dercole11, dercole12} set the 
theoretical framework to describe the formation of a second generation of stars in GCs, by 
gas ejected by AGB and SAGB stars, diluted with pristine gas having the original chemistry. 
These studies outlined the importance of the role played by SAGBs: these stars evolve
rapidly, and their winds could give origin to the formation of a He--rich stellar component, 
whose presence is suggested by photometric investigations of some GCs.

The yields from SAGBs are therefore crucial for this study. The spectroscopic analysis 
of stars belonging to the blue MS of GCs constitute a valuable test of the self--enrichment 
by AGBs mechanism, because their surface chemistry should reflect the composition of the 
yields of SAGBs.

Our previous investigations on this topic were based on a single metallicity, Z=0.001, 
corresponding to the chemistry of GCs with intermediate metallicity \citep{vd09, vd11}. 
This limitation was caused by lack of SAGB models of different metallicity, with the 
exception of the compilation by \citet{siess10}. To allow a more complete analysis 
we present here AGB and SAGB models spanning the range of metallicities of GCs, ranging 
from the chemistry typical of low--metallicity GCs, $Z=3\times 10^{-4}$, to the more 
metallic clusters, i.e. $Z=8\times 10^{-3}$. 

Although we discuss the implications for the self--enrichment by AGB \& SAGB stars,
this investigation is focused on the properties of the models, while the analysis of
how these new yields compare with the observations of the GCs of the same metallicity
is postponed to future investigations.

\section{The models}
\label{models}
The models presented and discussed in this paper were calculated by means of the
ATON code for stellar evolution. A detailed description of the numerical structure 
of the code can be found in \citet{ventura98}. 

The metallicities investigated are $Z=8\times 10^{-3}$ and $Z=3\times 10^{-4}$. 
The mixtures follow the relative abundances of the elements according to \citet{gs98}, 
with $\alpha-$enhancement $[\alpha/$Fe$]=+0.2$ (for $Z=8\times 10^{-3}$) and
$[\alpha/$Fe$]=+0.4$ (for $Z=3\times 10^{-4}$). These choices correspond to
[Fe/H]$=-0.5$ and [Fe/H]$=-2$, typical of high metallicity GCs, such as NGC 6388,
and low--metallicity structures, such as NGC 2419.

Convection was modelled in all cases by means of the Full Spectrum of Turbulence
(FST) model, presented by \citet{cm91}. The impact of this choice has been discussed
extensively in the literature \citep[e.g.][]{vd05}, and will not be repeated here.

In convectively unstable regions nuclear burning and mixing of chemicals were coupled
by means of a diffusive approach, using the scheme by \citet{ce76}. Overshoot of convective
eddies into radiatively stable regions is modelled by assuming an exponential decay
of convective velocities starting from the convective borders; the e--folding distance of
this behavior is given by $\sim \zeta H_P$, where $\zeta$ is the free parameter measuring
the extent of the extra--mixing. During the two main phases of core burning we assumed
$\zeta=0.02$, in agreement with \citet{ventura98}. We further explore the effects of a tiny 
extra--mixing from the convective shell that forms during the ignition of each
thermal pulse (Pulse Driven Convective Shell, hereianfter PDCS) by comparing the results 
obtained by neglecting any extra--mixing during the AGB evolution, with those found by 
assuming $\zeta=0.001$: this assumption was found to enhance the strength of thermal pulses,
and to an increase in the inwards penetration of the external mantle during the post--pulse
phases \citep{herwig00}. No overshoot was assumed from the bottom of the convective envelope 
during the AGB phase. These choices, far from being a true calibration of a still
largely unknown physical phenomenon, are intended to have a rough estimate on how the 
extent of the third dredge--up, and consequently the yields of the elements discussed in
this investigation, depend on the various factors that favor an enhancement of the
inwards penetration of the convective envelope in the phases following the thermal pulse.
A much more detailed treatment would be required for those phenomena strongly dependent
on the details of such a penetration, such as the s--process enrichment.

Mass loss was modelled according to \citet{blocker95}. The free parameter entering this
recipe was set to $\eta_R=0.02$, in agreement with the calibration based on the 
luminosity function of lithium--rich stars in the Magellanic Clouds given by
\cite{ventura00}. The Bl\"ocker description is based on a steep dependence of $\dot M$ on
luminosity, $\dot M \sim L^{4.7}$, in agreement with dynamical models of AGB envelopes
by \citet{bowen88}. For M stars, with masses $M>4M_{\odot}$, evolving at large luminosities,
the mass loss rates used here are substantially larger than other treatments in the 
literature, such as the classic recipe by \citet{VW93}, or the empirical relation by
\citet{vanloon05}; this difference will result in a smaller number of thermal pulses.
For models of smaller mass, the comparison between the Bl\"ocker formula and the treatment
for C--star winds by \citet{wachter02} is mainly determined by the strong sensitivity
to the effective temperature ($\sim T_{eff}^{6.81}$) of this latter. 
In this mass interval the mass loss
rates used here are smaller, although a comparison with other investigations in the
literature is not trivial, because the effects of the mass loss treatment are strongly 
related to the way convection is modelled, which is relevant for the effective
temperature of the models.

Radiative opacities for temperatures above 10000K were computed by the OPAL release, in
the version documented by \citet{iglesias06}. In the low temperature regime we
use the AESOPUS tool by \citet{marigo09}, that allows to account for changes in the
surface chemistry determined by TDU and HBB. Though more time consuming, this choice is
mandatory for a reliable description of the evolutionary phases during the AGB evolution
that follow the carbon enrichment of the external layers, as discussed in details in
\citet{vm09, vm10}. The conductive opacities were taken from Poteckhin (2006)\footnote{See 
the web page www.ioffe.rssi.ru/astro/conduct/}, and are added harmonically to the 
radiative opacities. 

Tables of the equation of state are generated in the (gas) 
pressure-temperature plane, according to the latest release of the OPAL EOS (2005), 
overwritten in the pressure ionization regime by the EOS by \citet{saumon95}, and 
extended to the high-density, high--temperature domain according to the treatment by 
\citet{SB00}.

The relevant cross-sections are taken from the recommended values in the NACRE 
compilation \citep{angulo}, with only the following exceptions: 
$3\alpha$ \citep{fynbo}; $^{12}$C($\alpha$,$\gamma$)$^{16}$O \citep{kunz};
$^{14}$N(p,$\gamma$)$^{15}$O \citep{formicola};
$^{22}$Ne(p,$\gamma$)$^{23}$Na \citep{hale02}; $^{23}$Na(p,$\gamma$)$^{24}$Mg 
\citep{hale04}; $^{23}$Na(p,$\alpha$)$^{20}$Ne \citep{hale04}; 
$^{25}$Mg(p,$\gamma$)$^{26}$Al (NACRE, upper limits);
$^{26}$Mg(p,$\gamma$)$^{27}$Al (NACRE, upper limits).

\begin{figure*}
\begin{minipage}{0.45\textwidth}
\resizebox{1.\hsize}{!}{\includegraphics{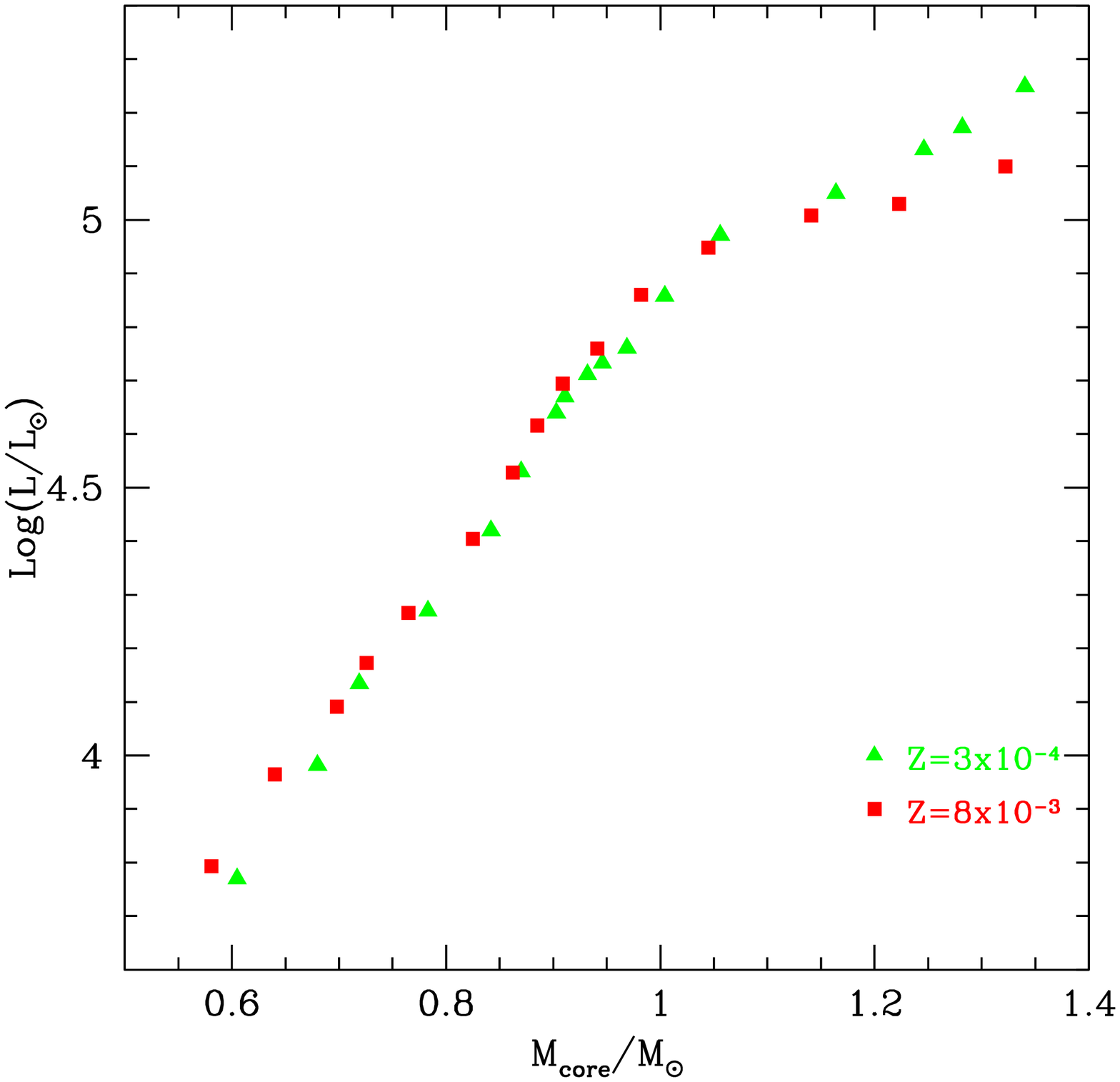}}
\end{minipage}
\begin{minipage}{0.45\textwidth}
\resizebox{1.\hsize}{!}{\includegraphics{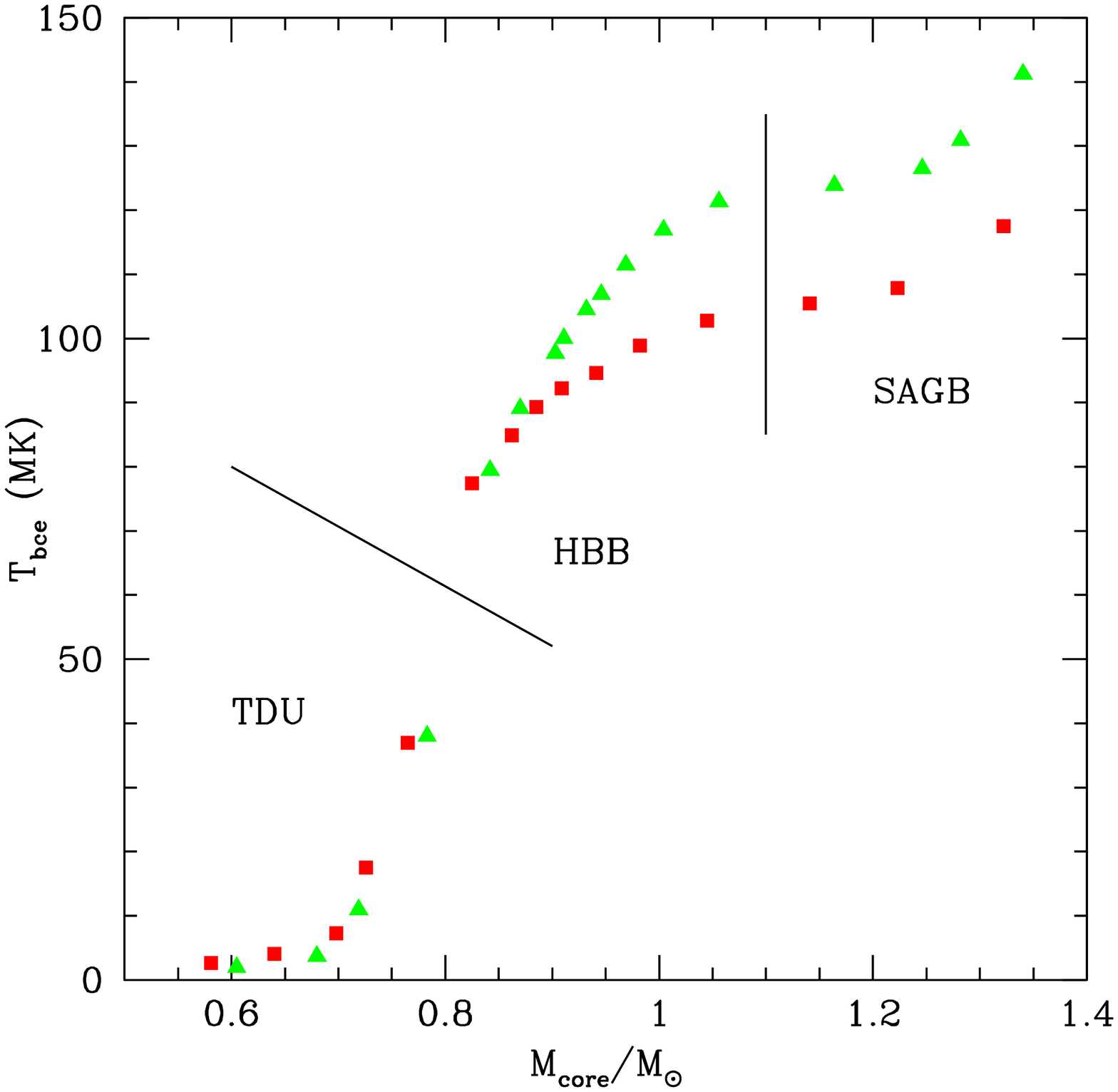}}
\end{minipage}
\caption{The maximum luminosity achieved during the AGB evolution (left), and the
maximum temperature reached at the bottom of the surface convective zone (right)
as a function of the core mass, for models with metallicity $Z=3\times 10^{-4}$
(blue triangles) and $Z=8\times 10^{-3}$ (red squares).
}
\label{physics}
\end{figure*}

\begin{table*}
\caption{Relevant properties of AGB models}
\label{yields}
\begin{tabular}{cccccccccccccccc}
\hline
\hline
M & $\tau_{\rm evol}$ & $N_p$ & $\tau_i$ & $M_c^{1TP}$ & $M_c$ & $T_{\rm bce}^{\rm max}$ & 
Y & Li & C & N & O & Na & Mg & Al & Si  \\
\hline
& & & & & & & $Z=3\times 10^{-4}$ & & & & & & & & \\
\hline
1.5  & 1.06e9  & 13 & 1.5e5  & 0.58 &  0.68  &  4e6     &  0.27  &  -1.03  &   1.99  &  0.27  &   0.69  &  0.19  &  0.42  &  0.12  &  0.40   \\
2.0  & 8.05e8  & 21 & 1.0e5  & 0.59 &  0.72  & 1.1e7    &  0.28  &   1.06  &   2.56  &  0.59  &   1.31  &  0.86  &  0.78  &  1.00  &  0.40   \\
2.5  & 4.55e8  & 22 & 3.0e4  & 0.72 &  0.78  & 3.8e7    &  0.26  &   1.17  &   2.28  &  0.52  &   1.24  &  0.66  &  0.73  &  1.09  &  0.41   \\
3.0  & 2.94e8  & 27 & 1.3e4  & 0.80 &  0.84  & 7.9e7    &  0.27  &   2.82  &   1.35  &  2.59  &   1.09  &  1.62  &  0.61  &  0.93  &  0.41   \\
3.5  & 2.07e8  & 34 & 8.5e3  & 0.83 &  0.87  & 8.9e7    &  0.27  &   2.33  &   0.36  &  1.87  &   0.49  &  1.10  &  0.42  &  0.40  &  0.41   \\
4.0  & 1.55e8  & 41 & 5.5e3  & 0.86 &  0.90  & 9.8e7    &  0.29  &   2.06  &   0.12  &  1.76  &   0.19  &  0.91  &  0.37  &  0.76  &  0.41   \\
4.2  & 1.38e8  & 42 & 5.1e3  & 0.88 &  0.91  & 1.e8     &  0.33  &   2.27  &   0.35  &  2.10  &   0.09  &  0.83  &  0.25  &  1.12  &  0.42   \\
4.5  & 1.20e8  & 47 & 3.6e3  & 0.90 &  0.93  & 1.02e8   &  0.33  &   2.14  &   0.23  &  1.94  &  -0.13  &  0.61  &  0.07  &  1.16  &  0.46   \\
4.7  & 1.09e8  & 49 & 2.8e3  & 0.91 &  0.95  & 1.07e8   &  0.34  &   2.02  &   0.10  &  1.83  &  -0.31  &  0.41  & -0.05  &  1.21  &  0.53   \\
5.0  & 9.60e7  & 53 & 2.1e3  & 0.94 &  0.97  & 1.11e8   &  0.35  &   2.05  &   0.24  &  1.60  &  -0.50  &  0.19  & -0.18  &  1.04  &  0.61   \\
5.5  & 7.90e7  & 54 & 1.3e3  & 0.98 &  1.00  & 1.17e8   &  0.36  &   1.95  &   0.07  &  1.41  &  -0.62  &  0.03  & -0.13  &  0.89  &  0.63   \\
6.0  & 6.70e7  & 41 & 1.0e3  & 1.02 &  1.05  & 1.21e8   &  0.36  &   1.90  &  -0.51  &  1.40  &  -0.82  &  0.01  &  0.02  &  0.86  &  0.59   \\
6.5  & 5.70e7  & 39 & 4.5e2  & 1.14 &  1.16  & 1.24e8   &  0.36  &   1.58  &  -0.59  &  1.39  &  -0.77  & -0.04  &  0.07  &  0.86  &  0.57   \\
7.0  & 4.95e7  & 37 & 3.2e2  & 1.21 &  1.25  & 1.27e8   &  0.37  &   1.89  &  -0.50  &  1.44  &  -0.43  &  0.08  &  0.22  &  0.82  &  0.51   \\
7.2  & 4.70e7  & 35 & 2.9e2  & 1.25 &  1.28  & 1.31e8   &  0.37  &   2.17  &  -0.43  &  1.49  &  -0.23  &  0.29  &  0.27  &  0.75  &  0.49   \\
7.5  & 4.35e7  & 31 & 2.6e2  & 1.30 &  1.34  & 1.41e8   &  0.37  &   3.17  &  -0.27  &  1.57  &   0.07  &  0.81  &  0.35  &  0.60  &  0.44   \\
\hline
& & & & & & & Z=$10^{-3}$ & & & & & & & & \\
\hline
2.0  & 9.70e8  & 20 & 2.1e5  & 0.53 &  0.69  & 7e6      &  0.26  &   1.44   &   1.45  &  0.48  &   0.56  &  0.33  &  0.41  &  0.13   &  0.40   \\
2.5  & 5.35e8  & 21 & 6.7e4  & 0.64 &  0.74  & 1.8e7    &  0.25  &   0.44   &   1.68  &  0.51  &   0.98  &  0.39  &  0.49  &  0.59   &  0.40   \\
3.0  & 3.40e8  & 26 & 1.8e4  & 0.77 &  0.82  & 6.5e7    &  0.25  &   2.53   &   0.84  &  2.21  &   0.92  &  1.16  &  0.57  &  0.65   &  0.40   \\  
3.5  & 2.35e8  & 31 & 1.2e4  & 0.81 &  0.85  & 8.3e7    &  0.26  &   2.33   &   0.51  &  2.18  &   0.77  &  1.30  &  0.55  &  0.66   &  0.40   \\
4.0  & 1.74e8  & 34 & 7.9e3  & 0.84 &  0.88  & 8.9e7    &  0.28  &   2.06   &   0.14  &  2.02  &   0.44  &  1.18  &  0.48  &  0.55   &  0.40   \\
4.5  & 1.33e8  & 39 & 5.3e3  & 0.87 &  0.91  & 9.5e7    &  0.31  &   1.89   &   0.12  &  1.89  &   0.19  &  0.97  &  0.41  &  0.96   &  0.41   \\  
5.0  & 1.06e8  & 40 & 3.0e3  & 0.91 &  0.94  & 1.01e8   &  0.32  &   1.97   &   0.13  &  1.70  &  -0.06  &  0.60  &  0.35  &  1.02   &  0.43   \\
5.5  & 8.48e7  & 41 & 2.0e3  & 0.96 &  0.98  & 1.06e8   &  0.33  &   1.99   &  -0.41  &  1.51  &  -0.35  &  0.37  &  0.28  &  1.10   &  0.44   \\
6.0  & 7.11e7  & 34 & 1.2e3  & 1.00 &  1.03  & 1.12e8   &  0.34  &   2.18   &  -0.62  &  1.35  &  -0.40  &  0.31  &  0.27  &  1.04   &  0.45   \\
6.3  & 6.50e7  & 33 & 7.9e2  & 1.03 &  1.06  & 1.14e8   &  0.35  &   2.22   &  -0.68  &  1.33  &  -0.37  &  0.30  &  0.30  &  0.99   &  0.44   \\
6.5  & 6.07e7  & 32 & 5.0e2  & 1.10 &  1.12  & 1.16e8   &  0.35  &   2.36   &  -0.71  &  1.31  &  -0.24  &  0.32  &  0.23  &  0.80   &  0.44   \\
7.0  & 5.26e7  & 31 & 4.1e2  & 1.18 &  1.20  & 1.2e8    &  0.36  &   2.12   &  -0.69  &  1.31  &  -0.15  &  0.39  &  0.25  &  0.74   &  0.44   \\
7.5  & 4.62e7  & 29 & 2.9e2  & 1.25 &  1.28  & 1.27e8   &  0.36  &   2.75   &  -0.61  &  1.31  &   0.01  &  0.67  &  0.29  &  0.57   &  0.43   \\
8.0  & 4.18e7  & 28 & 2.1e2  & 1.32 &  1.34  & 1.36e8   &  0.35  &   4.39   &   0.23  &  1.31  &   0.20  &  1.00  &  0.29  &  0.40   &  0.42   \\
\hline
& & & & & & & Z=$8\times 10^{-3}$ & & & & & & & & \\
\hline
1.5  & 2.44e9  & 14 & 1.5e5  & 0.54 &  0.64  & 4e6      &  0.28  &  -2.73  &   0.36  &  0.01  &   0.21  &  0.10  &  0.20  &  0.00  &  0.20   \\
2.0  & 1.06e9  & 18 & 1.3e5  & 0.55 &  0.70  & 7e6      &  0.28  &  -2.51  &   0.72  &  0.01  &   0.22  &  0.16  &  0.21  &  0.00  &  0.20   \\
2.5  & 6.98e8  & 28 & 1.7e5  & 0.51 &  0.73  & 1.7e7    &  0.28  &   1.95  &   1.02  &  0.46  &   0.28  &  0.28  &  0.24  &  0.17  &  0.20   \\
3.0  & 4.13e8  & 28 & 7.0e4  & 0.62 &  0.76  & 3.7e7    &  0.28  &   1.31  &   0.86  &  0.42  &   0.23  &  0.25  &  0.23  &  0.18  &  0.20   \\
3.5  & 2.68e8  & 32 & 6.0e4  & 0.76 &  0.82  & 7.7e7    &  0.28  &   2.90  &  -0.32  &  1.11  &   0.18  &  0.85  &  0.22  &  0.16  &  0.20   \\
4.0  & 1.89e8  & 35 & 9.0e3  & 0.82 &  0.86  & 8.5e7    &  0.29  &   2.71  &  -0.44  &  1.06  &   0.07  &  0.81  &  0.20  &  0.15  &  0.20   \\
4.5  & 1.42e8  & 33 & 6.0e3  & 0.85 &  0.88  & 8.9e7    &  0.31  &   2.58  &  -0.97  &  1.05  &  -0.01  &  0.76  &  0.20  &  0.18  &  0.20   \\
5.0  & 1.11e8  & 37 & 4.0e3  & 0.88 &  0.91  & 9.2e7    &  0.32  &   2.35  &  -1.00  &  1.04  &  -0.05  &  0.71  &  0.19  &  0.23  &  0.20   \\
5.5  & 8.93e7  & 40 & 2.4e3  & 0.92 &  0.94  & 9.5e7    &  0.34  &   2.59  &  -0.83  &  1.03  &  -0.06  &  0.67  &  0.19  &  0.27  &  0.20   \\
6.0  & 7.38e7  & 29 & 1.5e3  & 0.96 &  0.98  & 9.9e7    &  0.35  &   2.63  &  -1.06  &  1.00  &  -0.03  &  0.65  &  0.19  &  0.25  &  0.20   \\
6.5  & 6.20e7  & 39 & 8.5e2  & 1.02 &  1.04  & 1.03e8   &  0.36  &   2.93  &  -0.95  &  0.99  &  -0.01  &  0.65  &  0.19  &  0.25  &  0.20   \\
7.0  & 5.32e7  & 24 & 6.0e2  & 1.12 &  1.14  & 1.05e8   &  0.36  &   3.27  &  -1.06  &  0.95  &   0.02  &  0.66  &  0.20  &  0.19  &  0.20   \\
7.5  & 4.64e7  & 21 & 4.2e2  & 1.20 &  1.22  & 1.08e8   &  0.37  &   2.98  &  -1.07  &  0.96  &   0.01  &  0.68  &  0.20  &  0.18  &  0.20   \\  
8.0  & 4.09e7  & 20 & 2.9e2  & 1.30 &  1.32  & 1.16e8   &  0.37  &   3.91  &  -1.08  &  0.93  &   0.04  &  0.74  &  0.19  &  0.18  &  0.20   \\
\hline
\end{tabular}
\end{table*}

\section{AGB and SAGB evolution: physical aspects}
The evolutions presented in this work have been followed from the pre-main sequence
throughout the AGB phase, until the almost complete ejection of the external envelope.
The evolution of models developing a degenerate core, and experiencing the helium flash, 
were stopped at the tip of RGB, and resumed with an artificial HB model, having the same 
core mass as the model at the RGB tip.
For $Z=3\times 10^{-4}$ the helium flash was experienced
by models with mass below 2M$_{\odot}$, whereas for $Z=8\times 10^{-3}$ the threshold
mass is $M<2.5M_{\odot}$.

Table 1 summarizes the main physical quantities of the models discussed here, and the
average chemistry of their ejecta. We also report the results for $Z=10^{-3}$ by
\citet{vd09, vd11}, and \citet{vm10}. The first seven columns 
contain information regarding the physical evolution of the models: 
initial mass (solar units), time after which the AGB phase begins (yr), number
of thermal pulses experienced, inter--pulse period (yr), core masses
at the first TP and at the maximum in luminosity (solar units), and maximum 
temperature reached by the bottom of the convective envelope. Cols. 8--16 give information 
on the average composition of the ejecta, that is the helium mass fraction, the lithium 
content (expressed in the standard notation for lithium, i.e. 
$\log \epsilon (Li)= \log (n(Li)/n(H))+12)$, and the abundances of the elements mostly
investigated in the spectroscopic surveys of GC stars, i.e. the CNO elements, sodium,
magnesium (where we intend the sum of the three isotopes), aluminium and silicon. The
abundances from carbon to silicon are given as [i/Fe]$=\log(X_i/X_{Fe})-\log(X_i/X_{Fe})_{\odot}$,
to allow a more straight comparison with the observations.

The two panels of Fig.~\ref{physics} show the main physical properties of the AGB evolution
of the models computed: in the left panel we show the highest luminosity reached 
during the AGB phase as a function of the corresponding core mass, whereas in the right
panel we report the highest temperature reached at the bottom of the convective mantle.

\begin{figure*}
\begin{minipage}{0.33\textwidth}
\resizebox{1.\hsize}{!}{\includegraphics{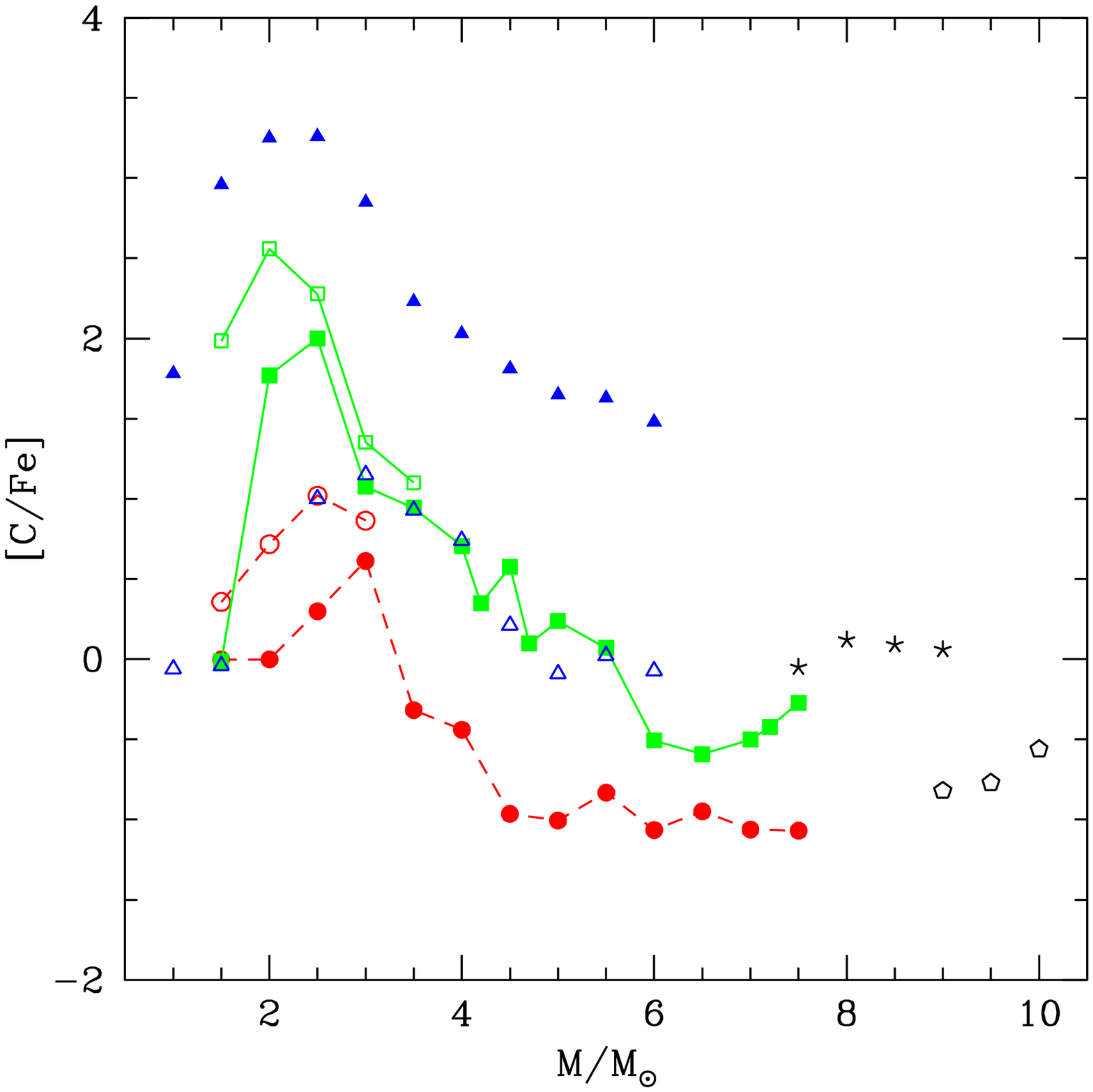}}
\end{minipage}
\begin{minipage}{0.33\textwidth}
\resizebox{1.\hsize}{!}{\includegraphics{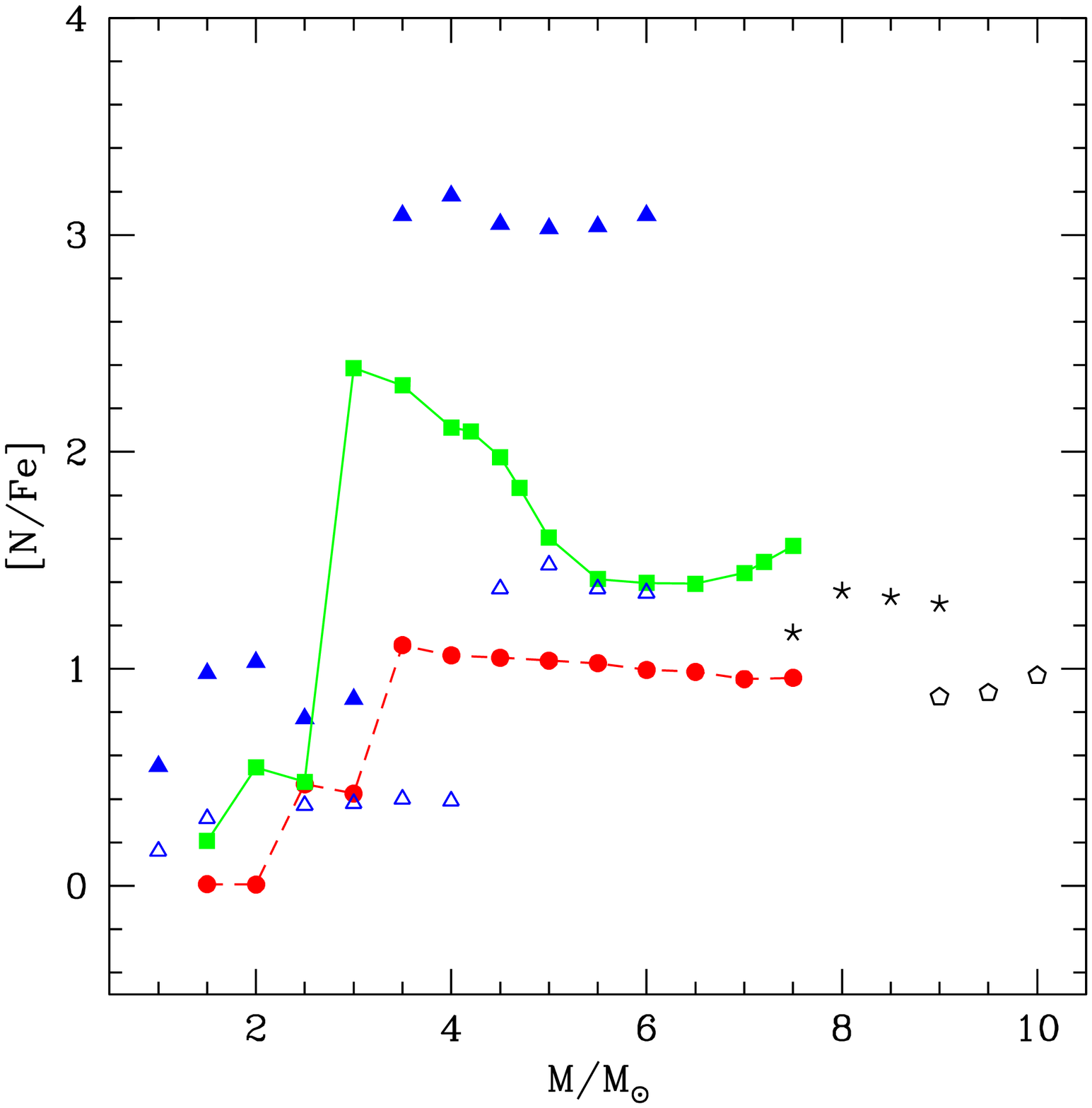}}
\end{minipage}
\begin{minipage}{0.33\textwidth}
\resizebox{1.\hsize}{!}{\includegraphics{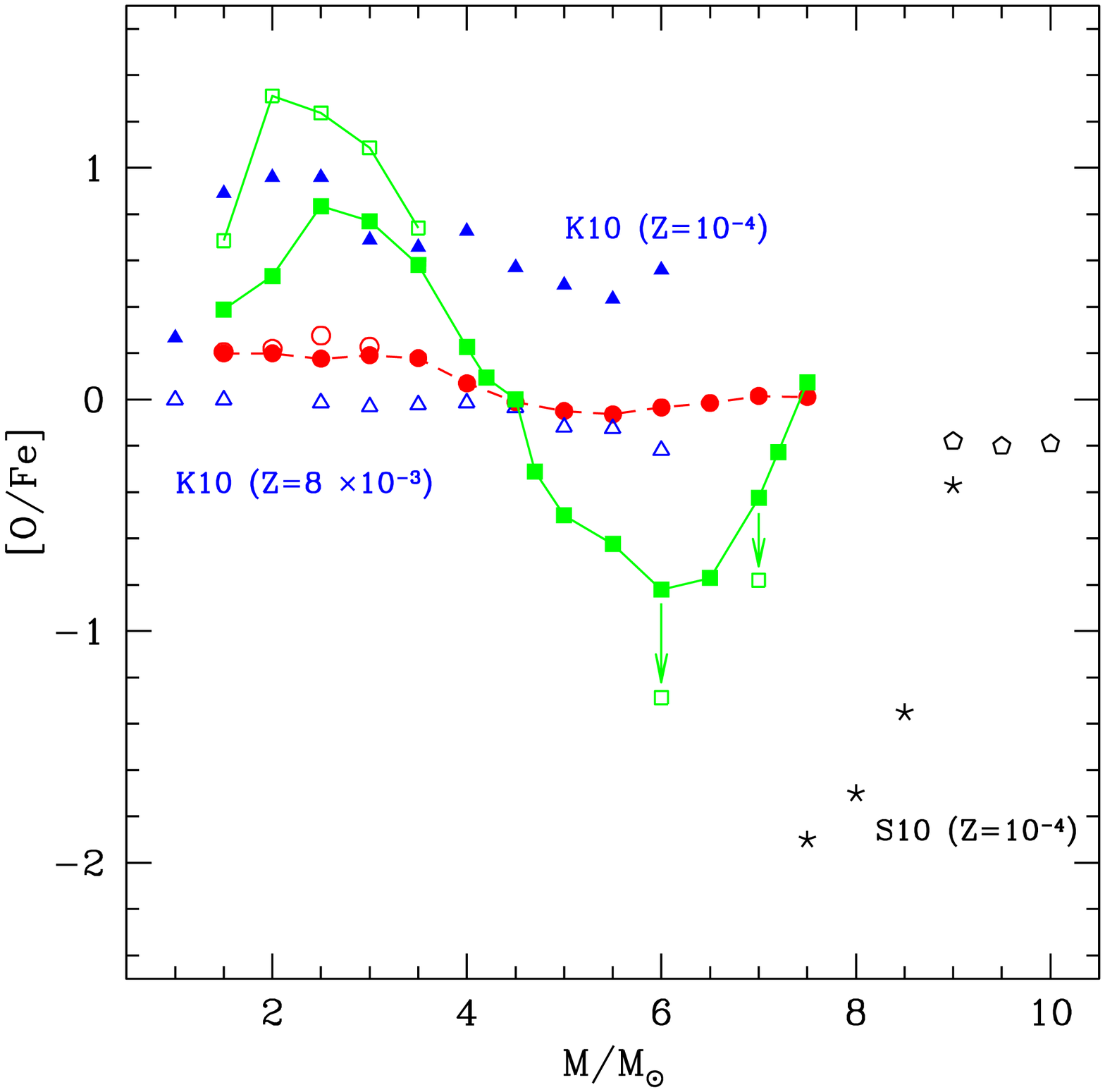}}
\end{minipage}
\caption{Average content of carbon (left panel), nitrogen (middle), oxygen
(right) for AGB and SAGB models of different metallicity, and with different
treatment of the borders of the convective shell that forms when the thermal pulses
occur. Models presented in this investigation with metallicity $Z=3\times 10^{-4}$ 
and $Z=8\times 10^{-3}$ are indicated, respectively in green (full squares) and
red (full dots); the same models, where some extra--mixing was assumed 
from the borders of the PDCS are indicated with open squares and dots. The Blue
points indicated models by \citet{karakas10} of metallicity $Z=10^{-4}$ (full 
triangles) and $Z=8\times 10^{-3}$ (open triangles). Black asterisks and open points
indicated the results by \citet{siess10} with chemistry, respectively, $Z=10^{-4}$
and $Z=8\times 10^{-3}$.
}
\label{cno}
\end{figure*}

The trend of luminosity with core mass is rather similar for the two metallicities, with
the exception of the models within the SAGB regime: the lower metallicity models 
reach larger luminosities.

In the right panel of Fig.~\ref{physics} we see a clear gap in temperature separating 
low--mass models, with core masses below $0.8M_{\odot}$, from their higher mass
counterparts. These latter experience Hot Bottom Burning (hereinafter HBB), consisting
in the nuclear activity at the bottom of the convective envelope when the temperature
exceeds $30-40$MK. HBB favours a rapid increase in the luminosity of the star
\citep{blocker91}, and its description depends on the details of convection modelling
\citep{renzini81, boothroyd88}. The threshold mass to
achieve HBB is 3M$_{\odot}$ and 3.5M$_{\odot}$, respectively, for the $Z=3\times 10^{-4}$
and $Z=8\times 10^{-3}$ cases. As will be discussed in more details in the following 
sections, HBB ignition has a strong influence on the chemical patterns. Models experiencing
HBB evolve at large luminosities, and loose their envelope much faster; compared to the lower 
masses, they undergo a limited number of thermal pulses, so that the effects of TDU in the 
alteration of the surface chemistry is modest \citep{vd08}.

Models whose core--mass exceeds $1.1M_{\odot}$ ignite carbon in an off--center, partially 
degenerate region, and evolve as SAGB stars.

\begin{figure*}
\begin{minipage}{0.45\textwidth}
\resizebox{1.\hsize}{!}{\includegraphics{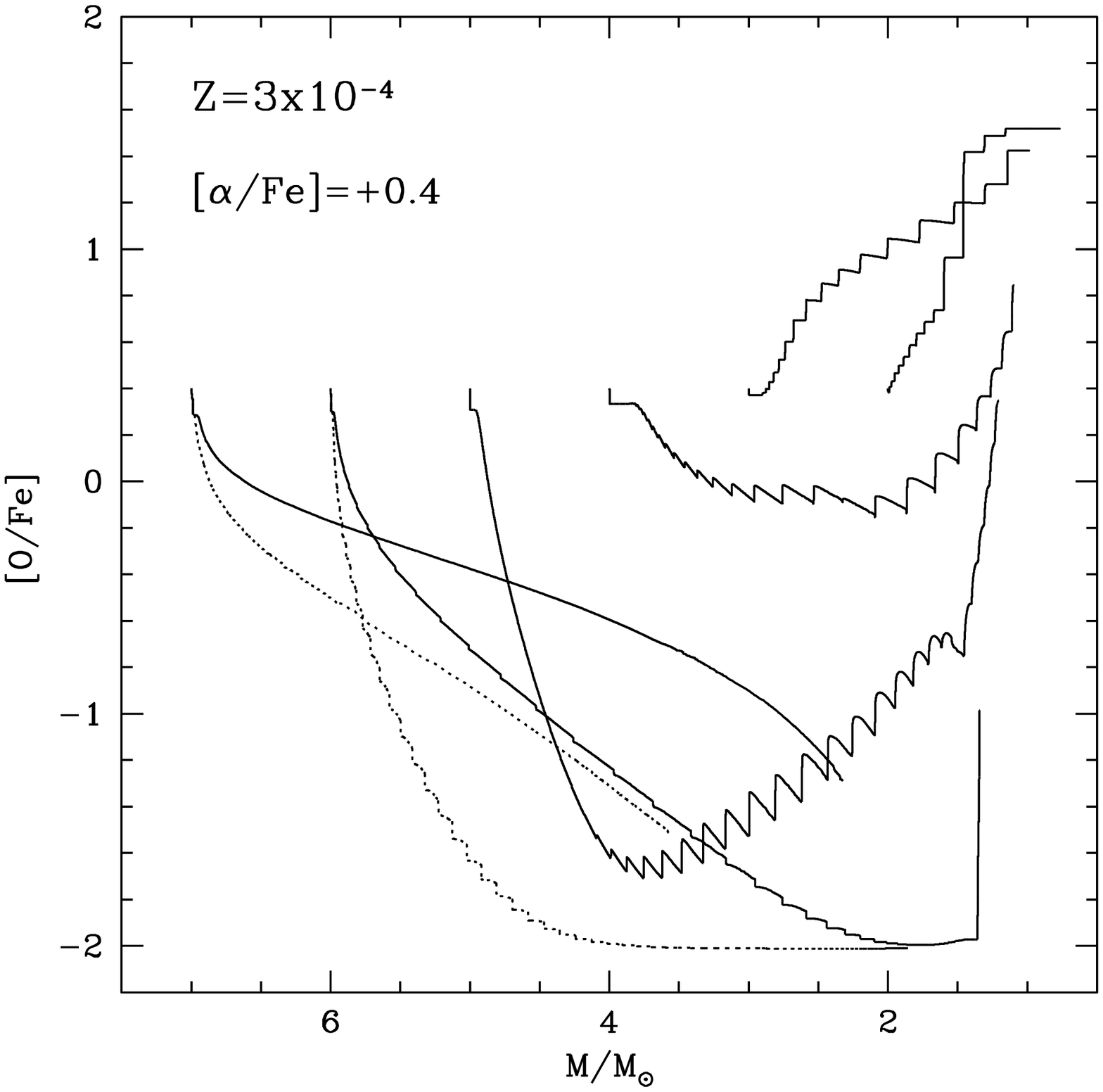}}
\end{minipage}
\begin{minipage}{0.45\textwidth}
\resizebox{1.\hsize}{!}{\includegraphics{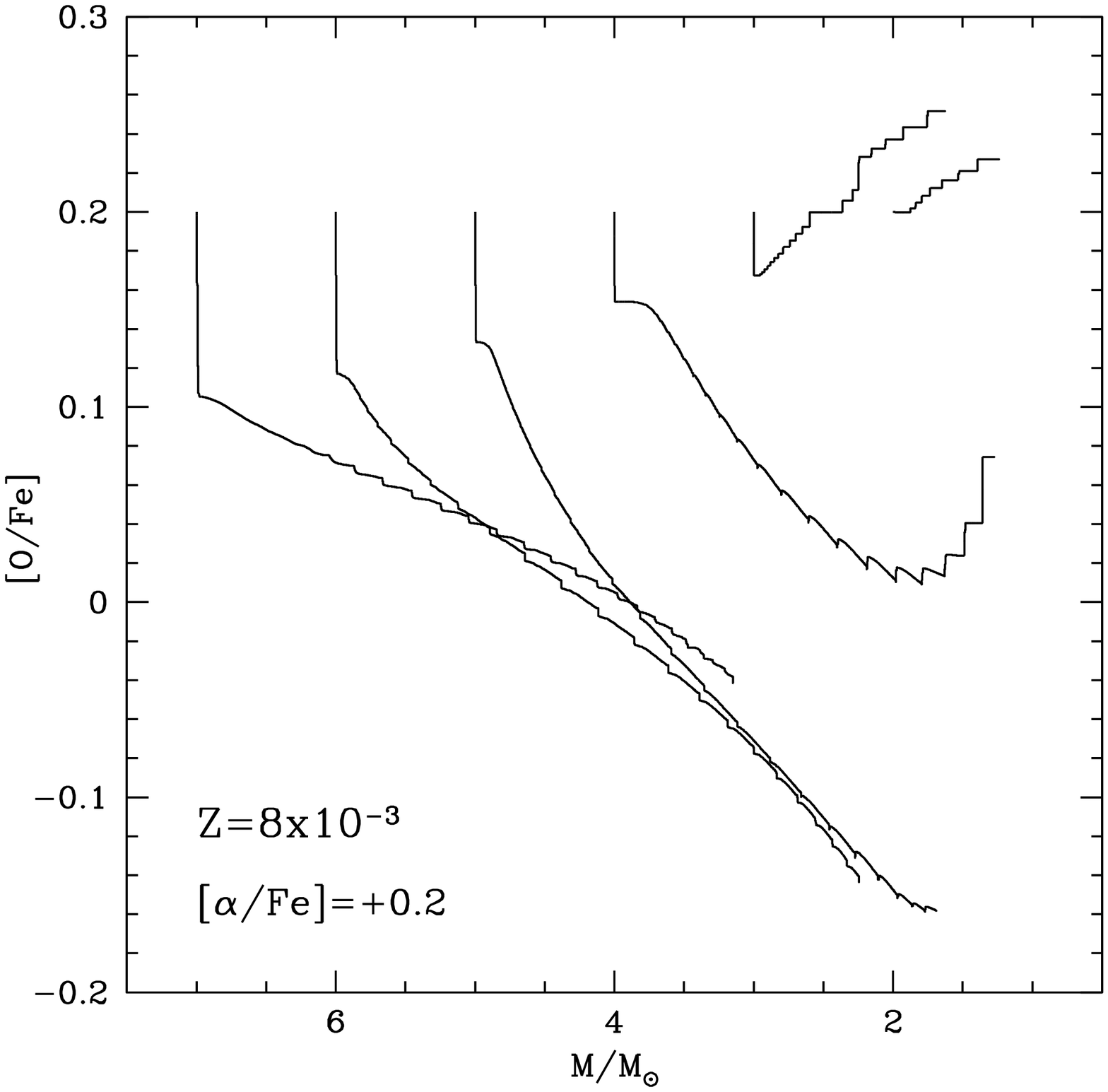}}
\end{minipage}
\caption{The variation of the surface oxygen during the AGB evolution of stars
of different mass, and metallicity $Z=3\times 10^{-4}$ (left panel) and
$Z=8\times 10^{-3}$ (right). The initial abundances of oxygen, in agreement with the
choice for the $\alpha-$enhancement of the two mixtures, are $[O/Fe]$=+0.4
($Z=3\times 10^{-4}$) and $[O/Fe]$=+0.2 ($Z=8\times 10^{-3}$). The effects of the
second dredge--up can be seen in the decrease in the oxygen content of the envelope at
the beginning of the AGB evolution. The two dotted lines in the left panel indicate
the results concerning two low--Z models with initial mass 6$M_{\odot}$ and 7$M_{\odot}$,
calculated with a smaller rate of mass loss. Use of different scales was made compulsory
by the different extent of both TDU and HBB for the two metallicities.
}
\label{o16}
\end{figure*}

The right panel of Fig.~\ref{physics} confirms that the strength of HBB is extremely
sensitive to metallicity. The $T_{bce}$ vs $M_{core}$ trend of the two sets of models
discussed here differ in the HBB domain. While in the $Z=8\times 10^{-3}$ models the
temperature at the bottom of the external mantle hardly reaches 100MK, in all 
low--Z models undergoing HBB we find $T_{bce} > 100MK$, with a maximum temperature,
approaching 150MK, reached inside the $7.5M_{\odot}$ model.
This is going to have a great impact on the extent of the nucleosynthesis experienced, 
given the extreme sensitivity to T of the cross--section of the various proton capture 
channels around $\sim 100$MK.

As discussed in section \ref{models}, the results obtained depend on the choices
made to model convection and mass loss. The treatment of convection is essential in
determining the strength of HBB: the range of masses experiencing HBB would be narrower
if a lower--efficiency convection model, such as the Mixing Length Theory, would be
used. Mass loss has no influence in determining whether HBB occurs for a given mass 
or not. However, changing the mass loss rate alters the rapidity with which mass loss
occurs, hence the number of TPs experienced. For models experiencing HBB, we have seen
that the Bl\"ocker's recipe used here leads to mass loss rates in excess with other
treatments in the literature: this favours a faster AGB evolution, and the contamination
of the surface chemistry determined by HBB will be softer, because there is no time of
achieving a very advanced nucleosynthesis. In the low--mass regime, for models reaching
the C--star stage, our recipe for mass loss leads to smaller rates: this reflects in a
difference in the final core mass of the models, which is $\sim 0.01-0.02M_{\odot}$ 
larger in the present case.

\section{Yields from AGBs and SAGBs}
The surface chemistry of AGBs is modified by TDU and HBB. The results obtained change
dramatically according to which of these two mechanisms is dominant, given the different 
chemical patterns produced. TDU determines an increase in the surface carbon, possibly
followed by nitrogen synthesis via proton capture during the quiescent phase of 
CNO burning. HBB changes the chemical composition according to the equilibrium abundances
associated to p--capture nucleosynthesis; this is extremely sensitive to the 
temperature at which HBB occurs. A fundamental difference between the effects of these
two mechanisms is that the overall C+N+O keeps constant as far as HBB dominates, whereas 
it increases if repeated TDU episodes occur.

\subsection{CNO}
The surface content of carbon, nitrogen and oxygen is touched by both TDU and HBB.
In the three panels of Fig.~\ref{cno} we show the average content of C (left), N (middle),
O (right) in the ejecta of stars with initial mass in the range 
$M_{\odot} \leq M \leq 8M_{\odot}$. To allow a more straight comparison with the 
spectroscopic analysis of stars in Globular Clusters, we show in the ordinate, for each
elements $i$, the quantity [i/Fe]. For a solar--scaled mixture, a positive [i/Fe]
indicates a production of the $i$--th element, whereas a negative value means that the 
element was destroyed. Note that for oxygen, which is an $\alpha$--element, the threshold 
values separating the production and destruction regimes are +0.2 and +0.4, respectively, 
for $Z=8\times 10^{-3}$ and $Z=3\times 10^{-4}$. The same holds for magnesium and silicon.

The $Z=8\times 10^{-3}$ models, indicated with red circles, are connected with dashed 
lines; the results for $Z=3\times 10^{-4}$ are indicated with squares, 
connected with solid lines. In the low--mass regime, full points indicate the 
results obtained by neglecting extra--mixing from the convective shell formed during 
the thermal pulses; models with some extra--mixing are indicated with open points.
Because this difference mainly reflects on the efficiency of the TDU, it has no influence
in the high--mass domain, where the chemistry is mostly determined by HBB.

[C/Fe] increases with mass in the low--mass domain (see left panel of Fig~\ref{cno}); 
this holds independently of metallicity, and is due to the higher number of thermal 
pulses experienced by higher mass models, allowing a greater enrichment in the surface 
carbon. The yields of lower Z models are predicted to be much richer in carbon,
because mixing is more efficient in low metallicity models \citep{boothroyd88}; also, 
for a given penetration of the surface convection determined by TDU, the percentage 
increase in the abundance of a given element is larger the lower is Z.

\begin{figure*}
\begin{minipage}{0.45\textwidth}
\resizebox{1.\hsize}{!}{\includegraphics{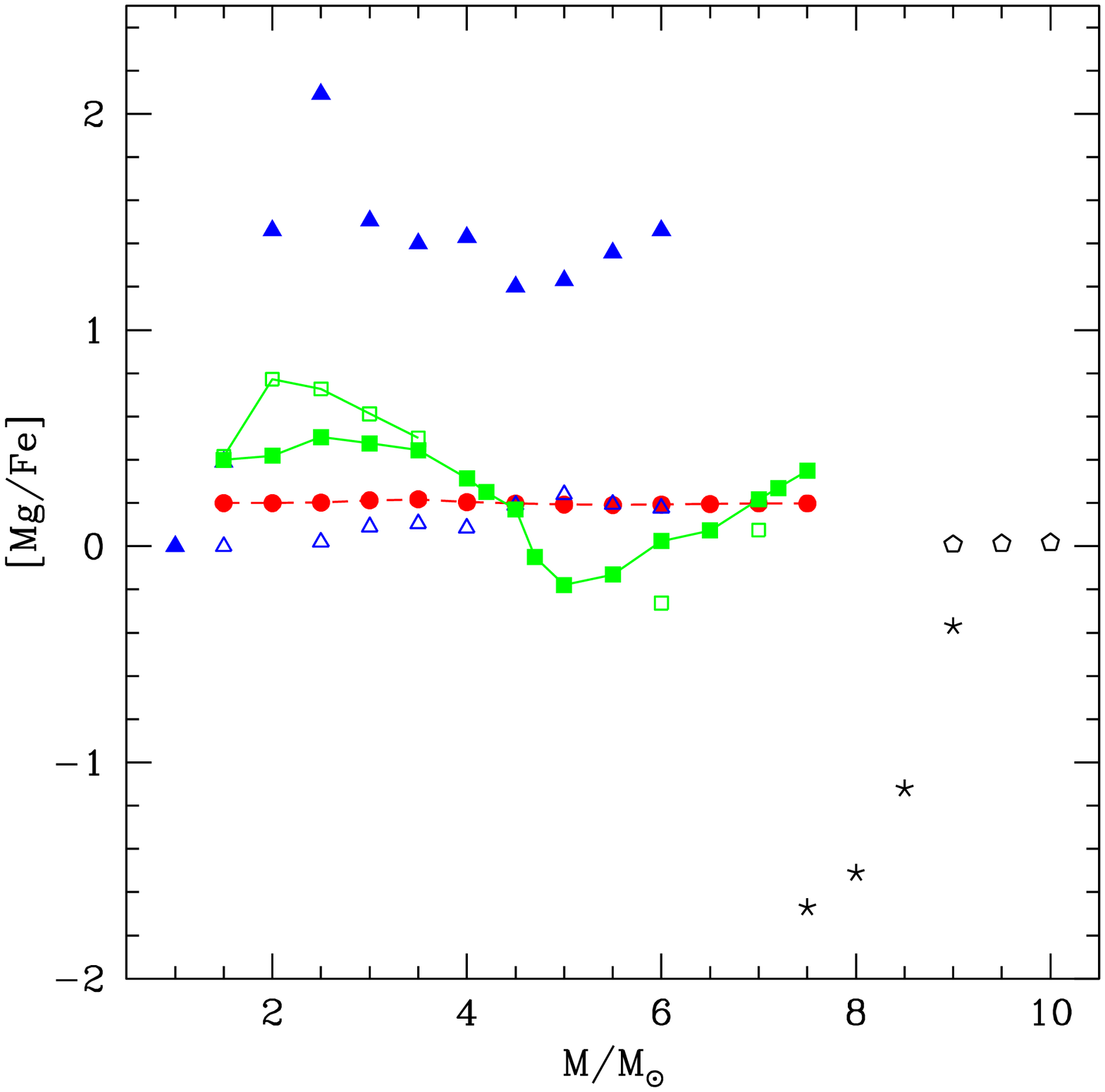}}
\end{minipage}
\begin{minipage}{0.45\textwidth}
\resizebox{1.\hsize}{!}{\includegraphics{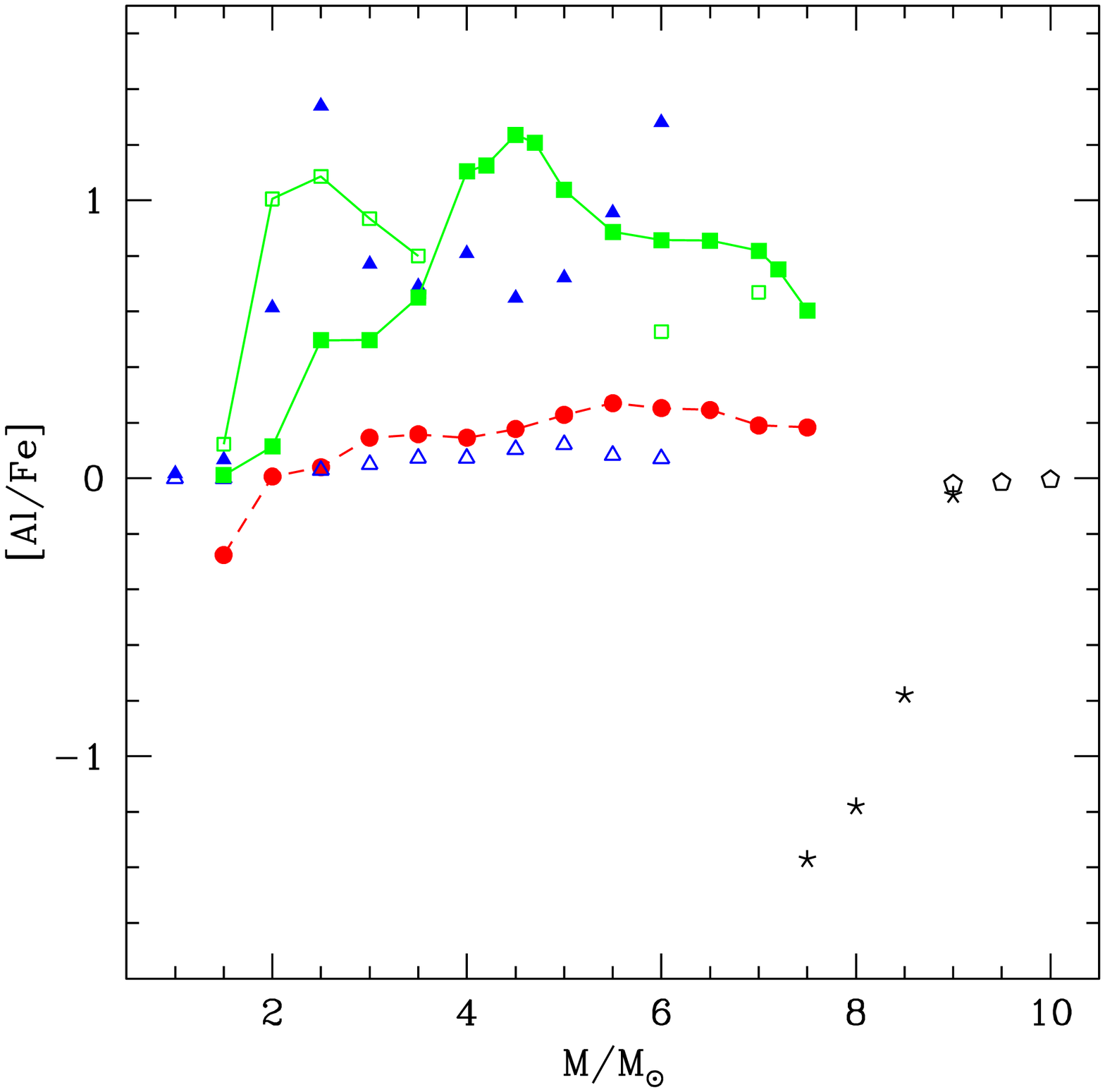}}
\end{minipage}
\caption{Average magnesium (left) and aluminium (right) content of the ejecta of models
presented in this investigation, compared to results by \citet{karakas10} and
\citet{siess10}. The meaning of symbols is the same as in Fig.~\ref{cno}.
}
\label{mgal}
\end{figure*}

Independently of the metallicity, assuming some extra--mixing from the borders of the
PDCS, in agreement with the results by \citet{herwig00}, favors the inwards penetration of
the base of the envelope in the post--pulse phase, thus leading to a more efficient TDU.
The carbon yields are consequently enhanced by a factor $\sim 10$.

In models experiencing HBB carbon is destroyed during the interpulse phase, which is
the reason for the negative trend with mass, shown in the left panel of Fig.~\ref{cno}. 
The $Z=3\times 10^{-4}$ models, though experiencing a stronger HBB (see right panel of
Fig.~\ref{physics}), show a higher carbon 
content, because of the effects of late TDU episodes, that increase the 
surface carbon content in the evolutionary phases preceding the White Dwarf 
cooling, just before the whole envelope is lost.
 
Nitrogen is not directly touched by TDU, which refurbishes the envelope of carbon and
(in minor quantities) of oxygen. This is the reason why models with mass below 
2--2.5M$_{\odot}$ show only a modest (if any) enhancement of nitrogen (see the middle panel
of Fig.~\ref{cno}). The average nitrogen in the ejecta shows a sharp increase for masses 
around the threshold value separating the TDU from the HBB regime: in these models,
the carbon carried to the surface by TDU is later converted into nitrogen during the
interpulse phase. 

Low Z models, experiencing a deeper TDU and a stronger HBB, are more efficient nitrogen
manufacturers. N is increased by a factor $\sim 300$ in the $Z=3\times 10^{-4}$ models
experiencing both TDU and HBB, and by a factor $\sim 15$ in $Z=8\times 10^{-3}$ models.

Higher mass models, experiencing only HBB, 
achieve a smaller production of nitrogen: in this case the increase is only due to the
conversion of the carbon and oxygen originally present in the gas, with no additional
contribution from TDU.

Unlike carbon, the predictions regarding the nitrogen content of the ejecta are 
more robust, and much less sensitive to the quantity of extra--mixing assumed from the 
borders of the PDCS. The influence of this latter is limited to the models achieving the
maximum production of nitrogen, and is at most a factor $\sim 2$ for masses 
$M\sim 2.5-3M_{\odot}$.

Oxygen is a key--element in the interpretation of the physical processes that alter the
surface chemical composition of AGBs: TDU produces a positive oxygen yield, whereas HBB
destroys it. The variation of the surface oxygen in some of the models discussed here is
shown in Fig.~\ref{o16}. The choice of the mass of the star (decreasing during the evolution)
as abscissa allows a better understanding of the average chemistry of the yields.

The solid and dashed lines in the right panel of Fig.~\ref{cno} (indicating,
respectively, the metallicities $Z=3\times 10^{-4}$ and $Z=8\times 10^{-3}$), 
trace similar patterns, with the difference that the lower--Z line is more 
stretched both upwards and downwards. Models with mass below $3.5-4M_{\odot}$ produce 
oxygen--rich ejecta (see Fig.~\ref{o16}), because TDU prevails over HBB. In analogy 
with carbon, we find that in the $Z=3\times 10^{-4}$ models the increase in the surface 
oxygen is larger. This can be clearly seen in the difference between the increase in [O/Fe]
found in the low--mass ($M \leq 3M_{\odot}$) models of $Z=3\times 10^{-4}$ (left panel of
Fig.~\ref{o16}), and the corresponding models of $Z=8\times 10^{-3}$ (right panel).

The oxygen yield becomes negative (i.e. [O/Fe] below +0.4 for 
$Z=3\times 10^{-4}$, and +0.2 for $Z=8\times 10^{-3}$) for $M\geq 4M_{\odot}$, where HBB 
destroys part of the surface oxygen. The effects of HBB can be seen in the decreasing
trend of the surface oxygen found for $M\geq 4M_{\odot}$ for both metallicities.

Because there is no way to produce oxygen efficiently via p--capture nucleosynthesis,
the history of oxygen under the effects of HBB is a pure destruction process, that 
proceeds at a higher rate for larger HBB temperatures. This motivates the smaller
oxygen in the ejecta of lower Z models, and also the decreasing trend with mass
for $M\geq 4M_{\odot}$ (see right panel of Fig.~\ref{cno}). The comparison among
lines corresponding to massive AGBs in the two panels of Fig.~\ref{o16} shows that
oxygen is destroyed much more strongly in the $Z=3\times 10^{-4}$ case (note the different
scales of the two panels).

The lowest oxygen abundances in the ejecta are found for $M \sim 6M_{\odot}$, with
[O/Fe]$=-0.8$ (note that considering the initial [O/Fe]$=+0.4$ for this mixture, this
corresponds to a reduction factor $\sim 20$). The trend with mass becomes increasing in
the SAGB regime, despite these models experience a stronger HBB (see right panel of
Fig.~\ref{cno}). This behaviour, discussed in \citet{vd11}, is due to the large mass
loss rates experienced by SAGBs, that loose their envelope before a great destruction
of the surface oxygen occurs. Both panels of Fig.~\ref{o16} show that the variation of
[O/Fe] with mass becomes less steep for $M \geq 6M_{\odot}$, confirming that the
mass loss rate of the largest masses is the key--factor for the larger oxygen in the 
ejecta of SAGBs.

That mass loss is the key--quantity in this context is 
confirmed by the results we obtain by assuming a smaller parameter entering the
Bl\"ocker's recipe, $\eta_R=0.005$ (corresponding to 1/4 of the standard value);
the corresponding oxygen abundances in the ejecta, indicated with open squares in 
Fig.~\ref{cno}, are consequently reduced. The results of these simulations are shown as
dotted lines in Fig.~\ref{o16}; we see that the smaller mass loss favors ejection of 
oxygen--poor gas, because little mass is lost in the early AGB phases, when the surface
oxygen is still large.

\subsection{The Mg-Al nucleosynthesis}
The surface abundances of magnesium and aluminium are altered by the effects of
TDU and HBB. The initial total magnesium in the stars is mainly under the form of 
$^{24}$Mg. In the 
regions touched by He--burning during the thermal pulses large amounts of
$^{25}$Mg and $^{26}$Mg are produced, via a series of $\alpha$--captures, that
start from $^{14}$N, and produce magnesium via $^{22}$Ne. 

The effects of HBB on the magnesium abundance are more complex, due to the various
proton capture reactions involving the different isotopes, ending up with the
synthesis of aluminium \citep{sa06, izzard07}. To destroy magnesium via HBB,
temperatures at the bottom of the convective zone of the order of $T\sim 10^8$K
are needed: this is required to activate efficiently the p--capture by $^{24}$Mg
nuclei. The outcome of magnesium burning is sensitive to the HBB temperature.
When the temperature is sufficiently large, great quantities of aluminium are 
produced, with also the synthesis of some silicon; conversely, when the temperature
is only slightly exceeding $10^8$K, most of the magnesium is locked in the
$^{25}$Mg isotope, with only a modest decrease in the total magnesium \citep{vcd11}.

Based on the above arguments, we understand that TDU and HBB produce opposite
effects on the surface magnesium: while TDU increases the Mg content, HBB
favors Mg--depletion, provided that the HBB temperatures exceed $\sim 10^8$K.
Because the initial magnesium is much larger (by $\sim$ a factor 50) than aluminium, 
even a weak magnesium burning is sufficient to increase the surface aluminium. 

Fig.~\ref{mgal} shows the magnesium (left panel) and aluminium (right) content of the
ejecta of the AGB models presented in this investigation. The trend of [Mg/Fe] with
mass is similar to oxygen. TDU in low--mass models favours an increase in the
$^{25}$Mg and $^{26}$Mg produced in the TPDS by the chain of $\alpha$--captures
mentioned above. In high--Z models the increase in the total magnesium is negligible, 
given the weaker efficiency of the TDU, and the higher initial magnesium in the envelope. 
In the low--metallicity case the increase in the surface total magnesium is sensitive to 
the efficiency of the TDU, as confirmed by the difference between the lines connecting open 
and full squares in the left panel, indicating, respectively, the results obtained with 
and without extra--mixing from the borders of the PDCS.

\begin{figure}
\begin{minipage}{0.45\textwidth}
\resizebox{1.\hsize}{!}{\includegraphics{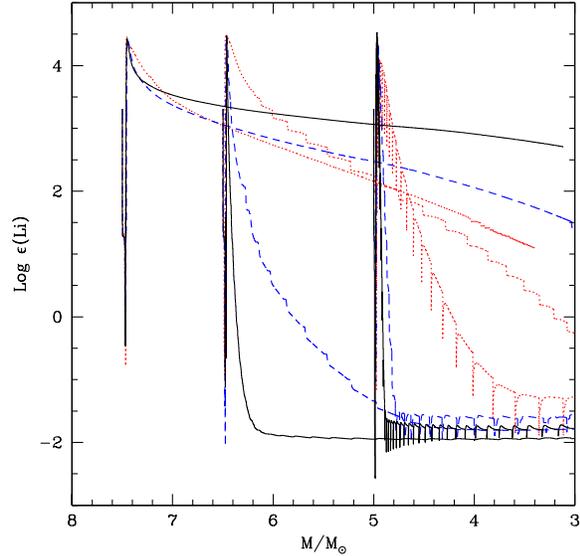}}
\end{minipage}
\caption{The variation during the AGB evolution of the surface lithium in models of
initial mass $5M_{\odot}$, $6.5M_{\odot}$, and $7.5M_{\odot}$, with metallicity
$Z=3\times 10^{-4}$ (black, solid line), $Z=10^{-3}$ (blue, dashed), $Z=8\times 10^{-3}$
(red, dotted).
}
\label{litio}
\end{figure}

The increase in magnesium favors a contemporary increase in the surface aluminium,
because part of $^{25}$Mg and $^{26}$Mg dredged--up to the surface is converted into 
aluminium during the following interpulse phase.

Models dominated by HBB show a different behavior: magnesium is destroyed by a series
of proton captures starting from the reaction $^{24}$Mg(p,$\gamma$)$^{25}$Al, that
eventually lead to the formation of aluminium (see e.g. \citet{vcd11} for the details
of the Mg--Al nucleosynthesis). The decrease in the total magnesium reaches a maximum
around $\sim 6M_{\odot}$ for the $Z=3\times 10^{-4}$ models, where the destruction
factor is $\sim 0.6$ dex; in the SAGB regime the magnesium yields are higher, because 
mass loss is so fast to prevent great destruction of magnesium before the envelope is
lost. Models achieving the greatest destruction of magnesium also produce
great quantities of aluminium, whose abundance is increased by a factor $\sim 20$
in comparison with the initial stellar content.

\subsection{Lithium}
AGB stars are known to evolve through a phase when they are efficient lithium manufacturers.
The mechanism by which lithium is produced in the envelope
of these stars was first suggested by \citet{cameron71}, and is activated whenever the
temperature at the bottom of the convective envelope reaches $\sim 40$MK. Under these
conditions $\alpha$--capture by $^3$He nuclei begins, with the production of beryllium,
which decays into lithium; because the time--scale for beryllium decay is $\sim 60$d, part of
the lithium is produced in the outermost and cooler regions of the envelope, where it
survives to proton fusion.

\citet{sackmann92} first found that lithium could be produced within the context of
AGB modeling, provided that a diffusive approach is used to couple nuclear burning and
mixing of chemicals in regions unstable to convection. \citet{mazzitelli99} confirmed
that lithium production can be activated efficiently when convection is modelled
according to the FST treatment, in all models with mass $M\geq 3M_{\odot}$. 

While the lithium--rich phase is crossed by all high--mass AGB models, the amount of
lithium which they eject into the interstellar medium is highly uncertain. 
The recent analysis by \citet{dantona09} outlines the various factors affecting the
lithium content of the AGBs ejecta. The uncertainty of the results obtained stems from
the fact that lithium production stops when the surface $^3$He is consumed: the quantity
of lithium expelled is determined by the mass lost by the star during the
phase when it is lithium--rich. 

The recent investigation by \citet{vd10} stressed that the lithium yields by SAGBs are
even more uncertain. When the same slope of the mass loss vs luminosity relation used
for AGBs is adopted, SAGB models are found to produce an extremely lithium--rich 
gas, because most of the mass is lost before $^3$He is consumed completely. However,
this latter finding is extremely sensitive to the mass loss rate adopted in the early
thermal pulses experienced, after the formation of the ONe core.

\begin{figure*}
\begin{minipage}{0.45\textwidth}
\resizebox{1.\hsize}{!}{\includegraphics{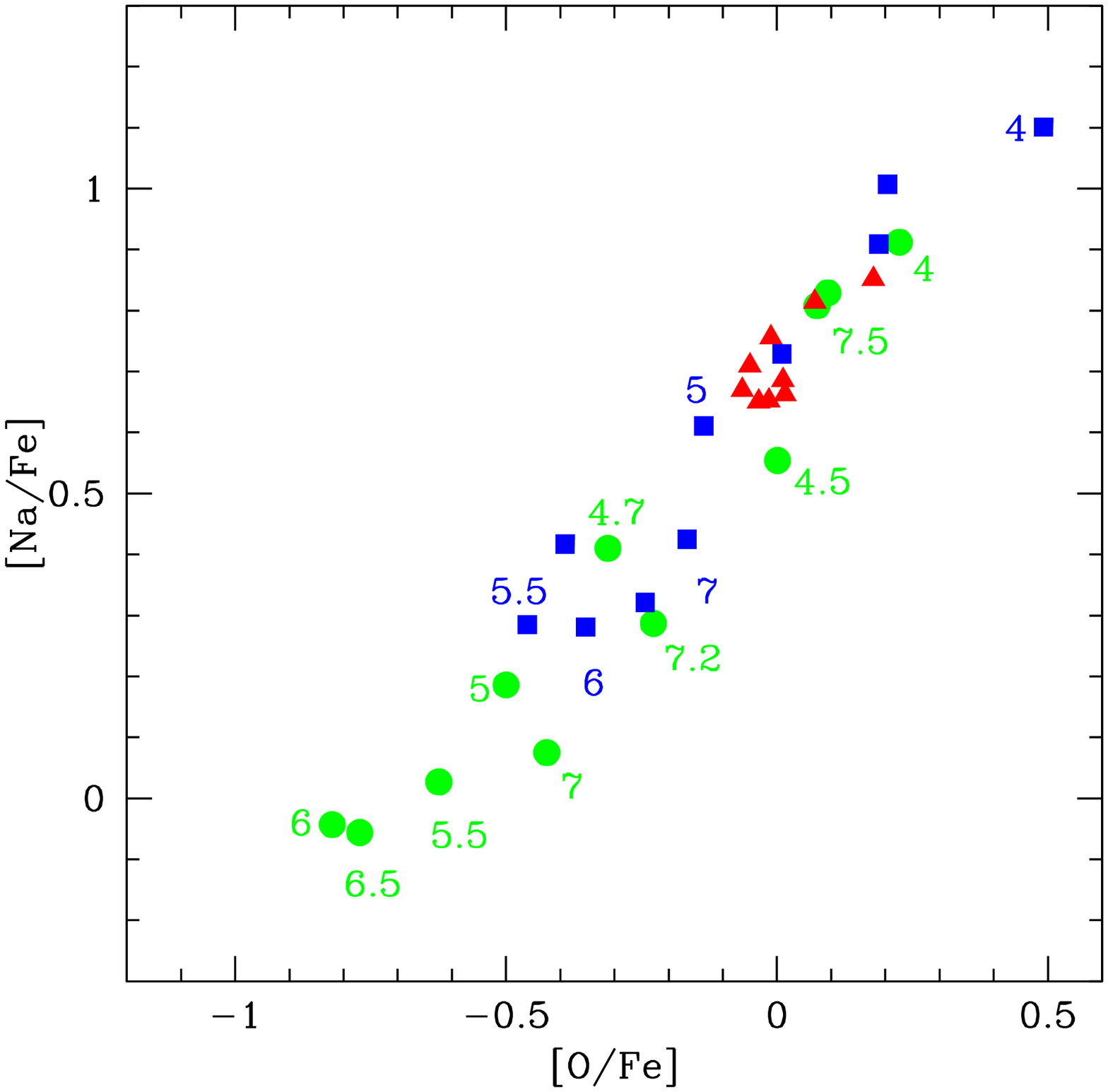}}
\end{minipage}
\begin{minipage}{0.45\textwidth}
\resizebox{1.\hsize}{!}{\includegraphics{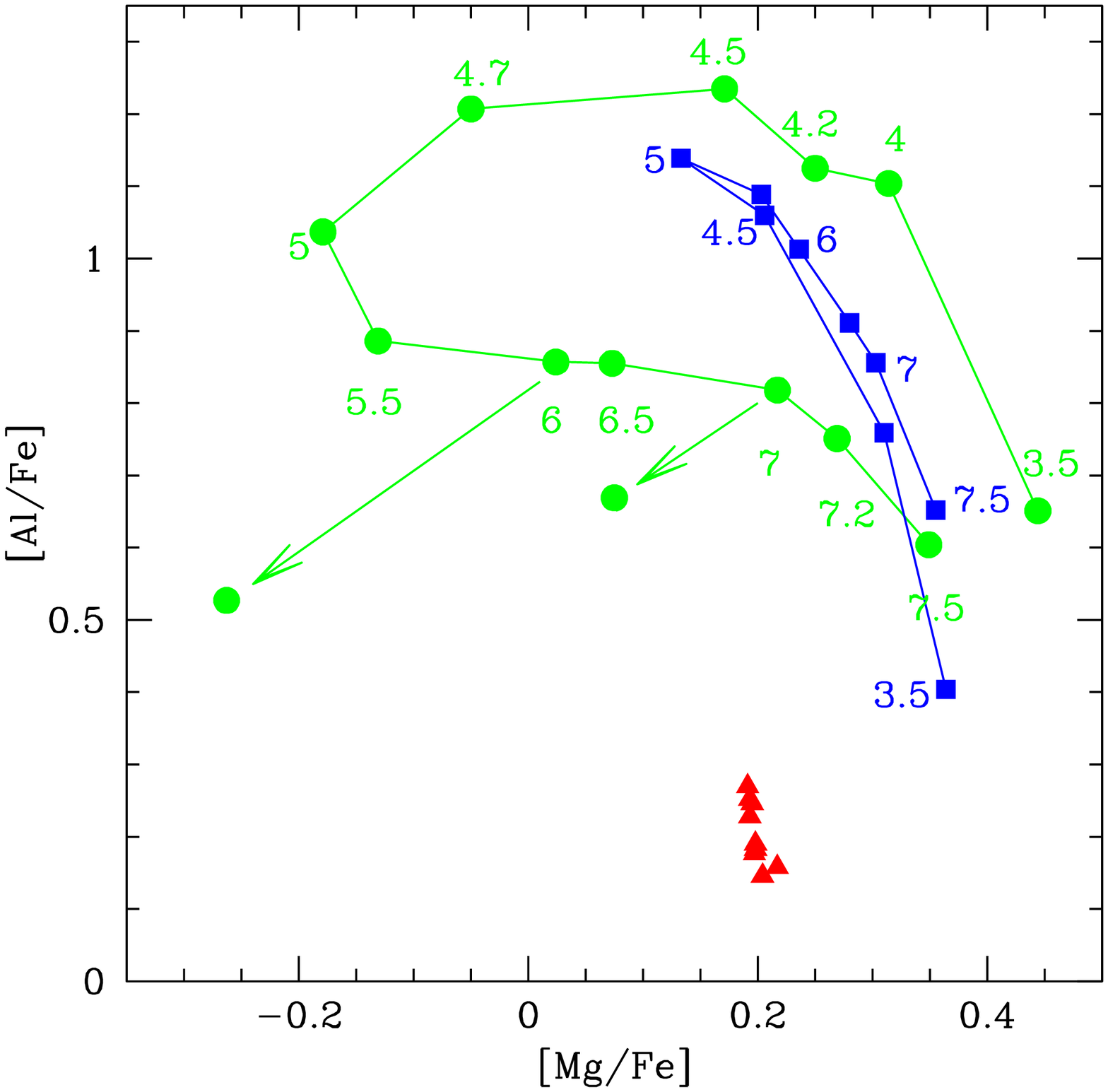}}
\end{minipage}
\caption{Yields of models of different metallicities in the O--Na and Mg--Al planes. 
Green points: $Z=3\times 10^{-4}$; blue squares: $Z=10^{-3}$ models by \citet{vd09, vd11};
red triangles: $Z=8\times 10^{-3}$. Numbers close to the points indicate the values of
the initial mass. The two arrows in the right panel indicate the results from $Z=3\times 10^{-4}$
models of initial mass $6M_{\odot}$ and $7M_{\odot}$ calculated with a smaller rate of mass loss.
}
\label{anticor}
\end{figure*}

The models presented in \citet{vd10} are compared with those in the present investigation in
Fig.~\ref{litio}. For clarity sake, we only show three models of initial mass
$5M_{\odot}$, $6.5M_{\odot}$, and $7.5M_{\odot}$. 

Lithium production is achieved by all the masses shown in Fig.~\ref{litio}, for the
three metallicities investigated. For the models of $5M_{\odot}$ and $6.5M_{\odot}$
the lithium content expected in the ejecta is directly correlated to metallicity:
higher Z models, for a given mass, experience a softer HBB, thus $^3$He is consumed more
slowly, and the duration of the lithium--rich phase is longer. 

In the range of mass $4M_{\odot} \leq M \leq 6M_{\odot}$ (see Table 1) the average lithium 
in the ejecta is $\log(\epsilon(Li)) \sim 2$ for $Z=3\times 10^{-4}$ and $Z=10^{-3}$, 
increasing to $\log(\epsilon(Li)) \sim 2.5$ for $Z=8\times 10^{-3}$. 

Moving from the AGB to the SAGB regime we see that the lithium yields increase with mass, as
also evident from Fig.~\ref{litio}: this is consistent with the arguments given above.

\subsection{Which implications for the self--enrichment of Globular Clusters?}
Massive AGBs and SAGBs have been proposed as one of the possible polluters of the 
interstellar medium in Globular Clusters: from the gas ejected by these sources
new stars would form, with the chemistry of their
ejecta. This is the reason why it is extremely important to understand which are the
predicted yields of these stars in terms of the elements commonly investigated in the
spectroscopic surveys of GC stars, i.e. oxygen, sodium, magnesium, aluminium and silicon.

The O--Na anticorrelation is
a common feature of all the galactic GCs so far examined, though the extension of the
observed pattern differs from cluster to cluster. The Mg--Al trend is also a rather
common feature, although a smaller amount of data are available, and in a few clusters 
the trend itself is not completely clear \citep{carretta09}.

The yields of the models presented in this investigation are shown in Fig.~\ref{anticor}, 
in the O--Na (left) and Mg--Al plane (right). We also show, for completeness, the results 
corresponding to $Z=10^{-3}$ from \citet{vd09} and \citet{vd11}.
The O--Na trend traced by the models confirm, on qualitative grounds, the main results 
outlined in \citet{vd11} (see their Fig.~6):

\begin{itemize} 

\item{The most extreme yields, i.e. those showing the greatest depletion of oxygen, 
are found for the masses at the edge between the AGB and the SAGB regimes, i.e. for 
M$\sim 6M_{\odot}$. The maximum extent of the oxygen destruction is extremely
sensitive to metallicity, ranging from $\delta [O/Fe] \sim -0.3$ for $Z=8\times 10^{-3}$,
to $\delta [O/Fe] \sim -0.8$ for $Z=10^{-3}$, up to $\delta [O/Fe] \sim -1.3$ for 
$Z=3\times 10^{-4}$.}

\item{Oxygen and sodium are correlated in all cases. In models where a strong destruction 
of oxygen occurs, the bottom of the surface convective zone is exposed to advanced p--capture
nucleosynthesis at temperatures $T > 100$MK: in this range of T's the destruction channel
for sodium is dominant compared to the production reaction by p--capture on $^{22}$Ne nuclei,
thus the sodium previously accumulated at the surface is destroyed. The oxygen yields
are negative, because HBB destroys oxygen; conversely, sodium can be produced, because 
of the initial increase in the surface sodium determined by the second dredge--up and
by the $^{22}$Ne burning via proton capture. Note that the positive correlation between 
oxygen and sodium is expected independently of all the uncertainties affecting the predictions 
concerning the sodium yield (initial neon and sodium abundances, cross--sections of the 
Ne--Na nucleosynthesis), that may eventually shift upwards or downwards the trend defined 
in Fig.~\ref{anticor}, without changing the slope.}

\end{itemize}

These results confirm that if massive AGBs were the polluters of the intra cluster 
medium in GCs, providing the gas from which new stellar generations formed, a certain
amount of dilution of the gas ejected via stellar winds with pristine, uncontaminated
matter is required, otherwise no O--Na anticorrelation can be produced.

Also, the GCs harboring a stellar generation formed directly from the ejecta of 
massive AGBs, with no dilution with pristine gas, must show a correlation of the
lowest oxygen abundances with metallicity, more metal--poor GCs showing the most
oxygen--poor population. In these stars we do not expect any sodium enhancement.

Turning to the Mg--Al cycling, we first note from the right panel of Fig.~\ref{anticor}
that only a modest reduction of the surface magnesium is achieved in all cases,
whereas some aluminium production, limited to +0.3 dex, occurs. This confirms that
magnesium destruction requires temperatures exceeding $\sim 10^8$K, only marginally
reached by the $Z=8\times 10^{-3}$ models (see right panel of Fig.~\ref{physics}).

Both the $Z=10^{-3}$ and the $Z=3\times 10^{-4}$ models produce magnesium--poor matter,
the maximum reduction factor being, respectively, $\delta$[Mg/Fe]$=-0.3$ and
$\delta$[Mg/Fe]$=-0.6$. Similarly to oxygen, we find that the most extreme chemistry
is not found for the most massive, SAGB models, because in the context of the present 
modeling these latter loose their envelope very rapidly, before a very advanced
Mg--Al nucleosynthesis may occur. 

The magnesium depletion is accompanied by an increase in the surface aluminium.
The Al content in the ejecta reaches a threshold value of [Al/Fe]$\sim 1.2$,
independently of the metallicity, and of the extent of the magnesium depletion. 
This is the effect of the balance reached between the production and destruction 
channels, that eventually leads to the formation of some silicon.

The uncertainties affecting the extent of magnesium depletion and aluminium enhancement
were discussed in details by \citet{vcd11}. The main outcome of this investigation is
that massive AGBs in the low--Z domain reach at the bottom of the surface convection zone
temperatures sufficiently large to destroy $^{24}$Mg, thus their ejecta are predicted to
show large $^{25}$Mg/$^{24}$Mg and $^{26}$Mg/$^{24}$Mg isotopic ratios. The maximum
depletion of the total magnesium was found to be determined by the cross section of the
$^{25}$Mg(p,$\gamma$)$^{26}$Al reaction in the range of temperatures around $\sim 10^8$K,
an increase by a factor 2 in the reaction rate corresponding to a further $\sim -0.2$~dex
in the magnesium depletion (see Fig.~5 in \citet{vcd11}). Even a larger Mg--depletion
would scarcely influence the Al--enhancement, due to the afore mentioned equilibrium
established between production and destruction rates, once Al increases by a
factor $\sim 10$.

\section{Comparison with other investigations in the literature}
The recent years have seen a growing interest towards AGB modeling. Several research groups
have contributed to produce an impressive series of models, spanning a wide range
of mass and metallicities. Some of these investigations were focused on the efficiency
of mixing in low--mass AGBs, and on the treatment of the borders at the convective/radiative
interface \citep{cristallo09, stancliffe07}. These studies complete the previous investigations 
on these evolutionary phases by \citet{kl03, kl07}, \citet{weiss09}, and the most recent 
update by \citet{karakas10}.

\begin{figure*}
\begin{minipage}{0.45\textwidth}
\resizebox{1.\hsize}{!}{\includegraphics{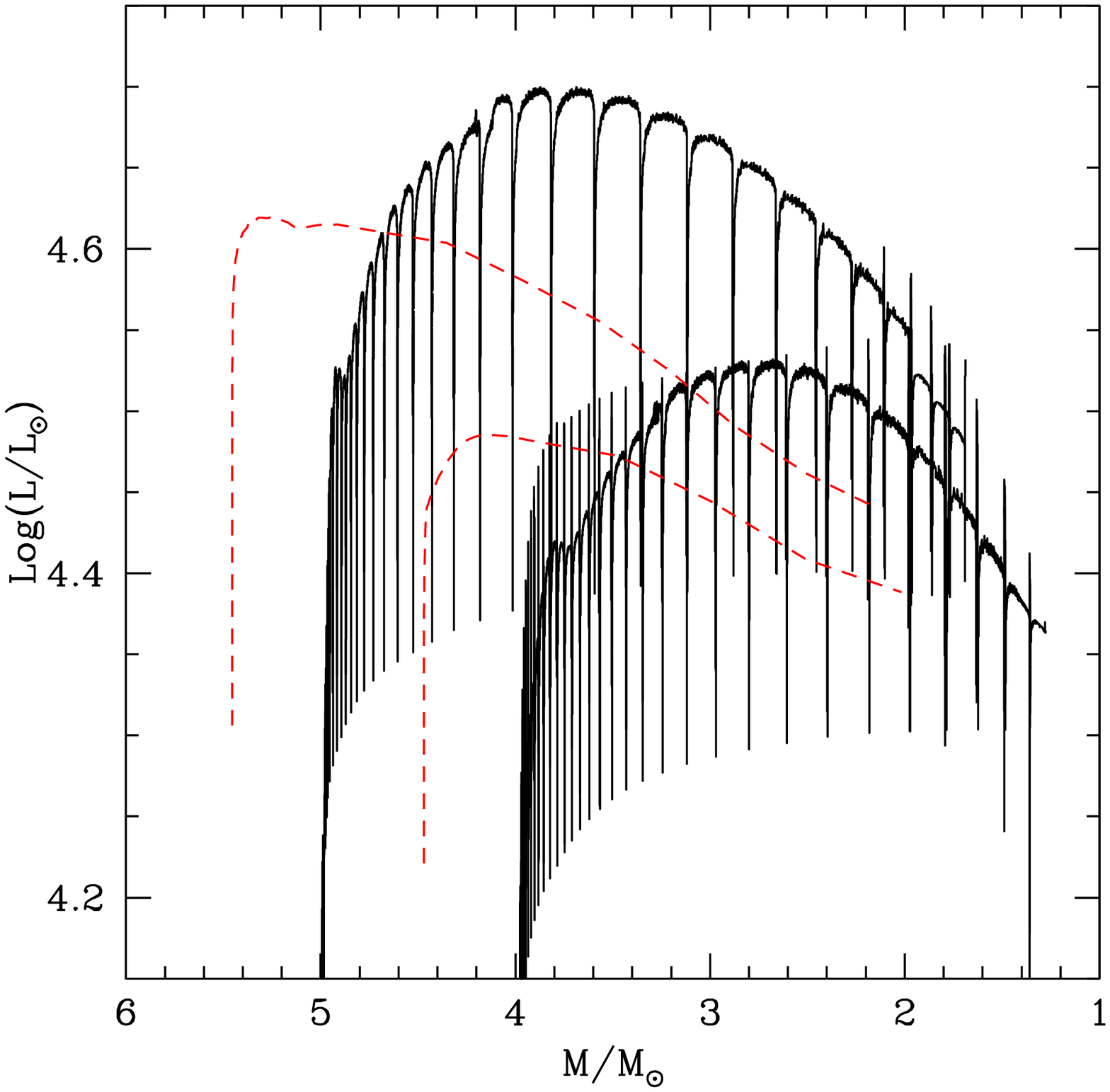}}
\end{minipage}
\begin{minipage}{0.45\textwidth}
\resizebox{1.\hsize}{!}{\includegraphics{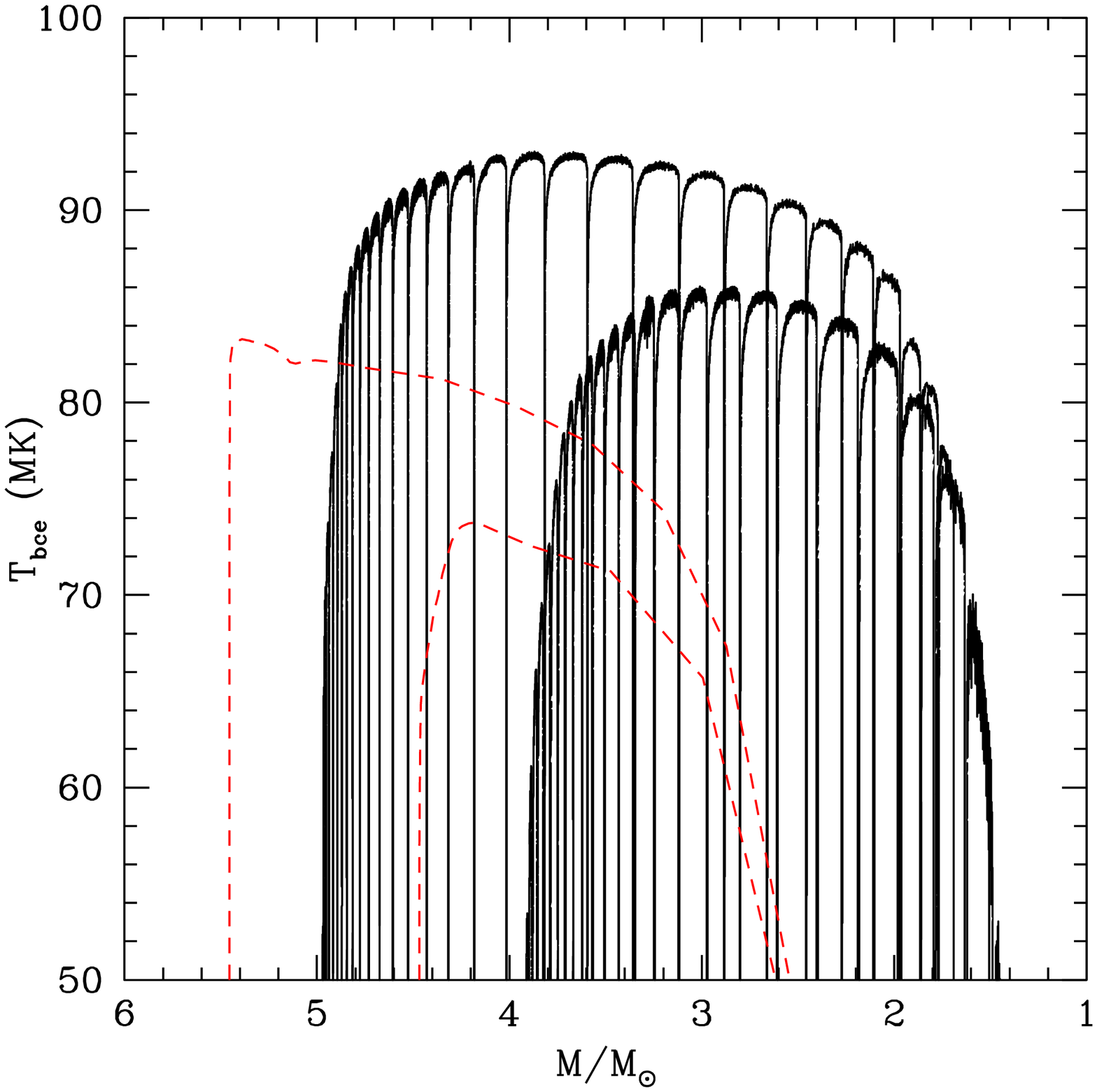}}
\end{minipage}
\caption{The evolution during the AGB phase of luminosity (left) and
temperature at the bottom of the envelope (right) of models from the
present investigation of initial mass $4M_{\odot}$ and $5M_{\odot}$
(black, solid tracks), and models by K10 with initial mass 
$4.5M_{\odot}$ and $5.5M_{\odot}$ (red, dashed lines). The initial masses
were chosen such that the models have the same core mass at the first 
thermal pulse.
}
\label{confz8m3}
\end{figure*}

On the side of SAGB modeling, the recent works by \citet{siess06, siess07, siess10}
provided a complete and exhaustive update of the pioneering explorations focused on
the physics of carbon ignition in regime of partial degeneracy, by \citet{ritossa96, ritossa99},
\citet{garcia97}, \citet{iben97}. 

Among the above investigations on the AGB phase, we decided to compare our results
with those by \citet{karakas10} (K10): in this compilation, the yields of
models with mass in the range $M_{\odot} \leq M \leq 6M_{\odot}$, and metallicities
in the range $0.0001 \leq Z \leq 0.02$ are presented and discussed. K10 yields
are indicated in Fig.~\ref{cno} and \ref{mgal} with blue triangles: the open points 
indicate the $Z=8\times 10^{-3}$ metallicity (to be compared with the red points in our 
compilation), whereas full points refer to $Z=10^{-4}$ (among the various 
metallicities treated in K10, this is the closest to the $Z=3\times 10^{-4}$ case 
discussed here).

As for the SAGB phase, the most complete investigation currently present in the
literature is by \citet{siess10} (hereinafter S10), which treats the same 
chemistry as K10, with masses in the range $7.5M_{\odot} \leq M \leq 10M_{\odot}$, 
i.e. those undergoing carbon ignition in conditions of partial degeneracy. The corresponding 
points are indicated as open pentagons ($Z=8\times 10^{-3}$) and asterisks
($Z=10^{-4}$) in Figs.~\ref{cno} and \ref{mgal}.

For what concerns carbon, for masses M$\leq 3M_{\odot}$, and metallicity 
$Z=8\times 10^{-3}$, our yields for models calculated with some extra--mixing from the 
PDCS are rather similar to those by \citet{karakas10}. The reason for this is in the
similarity of the physical properties of the two sets of models in this range of mass. A 
comparison with the models presented in \citet{karakas02} shows that for any given initial
mass the core masses at the first TP are very close, whereas the minimum core mass at which
TDU occurs is $\sim 0.02M_{\odot}$ larger in our case. The investigations by \citet{izzard04} 
and \citet{marigo07} showed that to reproduce the luminosity function of carbon stars in 
the Large Magellanic Cloud, whose metallicity is similar to $Z=8\times 10^{-3}$, the core 
masses when the first TDU occurs must be $\sim 0.07-0.1M_{\odot}$ smaller than in 
\citet{karakas02}. This means that the carbon yields found in the low--mass models presented
here must be considered as lower limits.

In the low--metallicity case the K10 carbon yields are a factor $\sim 3$ higher than 
ours; certainly the smaller metallicity upon which the 
computations by K10 are based ($Z=10^{-4}$ vs $Z=3\times 10^{-4}$) plays a role here,
although other factors, such as a different efficiency of TDU, cannot be ruled out as
possible reasons for this difference. In both compilations we see a decrease with mass 
in the carbon content of the ejecta, although in the models presented here the carbon 
yields become negative (i.e. the average carbon content of the ejecta is smaller as it 
was initially), whereas in K10 models [C/Fe] hardly reaches 0, rather it remains above 2 
in the low--Z case. This finding outlines the main difference between the models presented 
here and those by K10, i.e. the extent of HBB, much stronger in our case, because of the 
large convective efficiency predicted by the FST treatment; K10 models are based on the 
MLT treatment with $\alpha=1.75$, which leads to a much weaker HBB \citep{vd05}.

Fig.~\ref{confz8m3} compares the evolution of our models with metallicity $Z=8\times 10^{-3}$
(black, solid lines) with those of the same metallicity by K10 (red, dashed tracks). The initial
masses are slightly different, to compare models with the same core mass at the beginning of
the AGB phase: our $4M_{\odot}$ and $5M_{\odot}$ cases correspond, respectively, to masses 
$4.5M_{\odot}$ and $5.5M_{\odot}$ in K10. The differences in the HBB experienced can be clearly
seen in the comparison of the temperatures at the base of the envelope (see right panel): models 
by K10 are $\sim 20MK$ cooler than ours. This is also related to the different 
luminosity of the models (see right panel of Fig.~\ref{confz8m3}). Our models, evolving at 
larger luminosities, also suffer a larger mass loss, thus experience a smaller number of thermal
pulses: we find a total of 35 and 37 TPs for $M=4M_{\odot}$ and 
$M=5M_{\odot}$, to be compared to 38 TPs experienced by the K10 $4.5M_{\odot}$ model, and 
56 TPs in the $5.5M_{\odot}$ case.

HBB also affects the oxygen yields, as can be seen in
the right panel of Fig.~\ref{cno}. In the $Z=8\times 10^{-3}$ case we see that our models 
show some depletion of oxygen, by $\sim 0.2-0.3$dex, not found in
the investigation by \citet{karakas10}. The most striking difference is however found in
the low--Z case: while in the models presented here the ejecta are predicted to be 
extremely poor in oxygen, with a maximum depletion factor of the order of $\sim 30$,
in the K10 study we obtain, independently of the mass, oxygen--enriched 
matter, with [O/Fe]$>+0.4$. To understand this difference we compared in
details the most massive model in the K10 $Z=10^{-4}$ set, i.e. $M=6M_{\odot}$, with
our $Z=3\times 10^{-4}$ model of the same core mass, i.e. $M=5.5M_{\odot}$. As in the
previous comparison at larger metallicities, we find the same difference of $\sim 20MK$
(100MK in K10, and 120MK in our case) between the maximum temperatures experienced at
the bottom of the convective envelope. These different temperatures would explain the 
larger depletion of oxygen in our exploration, but not the large oxygen content in 
the K10 yields; this latter is due to the relevant contribution from TDU, partly
due to the smaller metallicity of K10 models, and also to the much larger number of
TPs experienced (the total number of TPs in K10 and our model are, respectively,
109 and 54).

The differences outlined above have also a feedback on the amount of nitrogen
that these stars produce. Upon discussing the CNO yields, we stressed that the most
efficient production of nitrogen takes place when TDU and HBB are both operating during
the AGB evolution, because carbon transported outwards by TDU is later converted 
into nitrogen. While in our case this overproduction of nitrogen is restricted to masses
$M\sim 2.5-3M_{\odot}$ (more massive stars experience only a small number of thermal 
pulses, due to the strong HBB), in K10 this behavior is shared
by practically all masses, as can be understood from the position of the full triangles 
in the middle panel of Fig.~\ref{cno}.

In the comparison with the work by \citet{siess10}, we preliminary note that the 
difference in the masses involved are due to the difference in the treatment of the
convective borders during the main sequence phase of hydrogen burning. We assumed some
overshoot from the border of the convective core, whereas S10 models were calculated with 
no extra--mixing: similar results are obtained for a mass $\sim 1.5-2M_{\odot}$ larger 
in Siess' computations, as can be seen in Fig.~\ref{cno}. 

Other than the shift in the initial mass, the $Z=8\times 10^{-3}$ models from the 
two compilations are fairly similar. In particular, we note that the maximum temperature
reached at the bottom of the envelope ranges between $\sim 10^8$K to $\sim 1.2\times 10^8$K 
(compare the values in the 7th col. of Tab.~\ref{yields} with those of col.15 in
\citet{siess10}). This is the reason for the similarity in the CNO yields at this metallicity. 
At these temperatures HBB is sufficiently strong to destroy carbon, and to synthesize nitrogen; 
oxygen is only marginally touched. 

Turning to the low--metallicity regime, the $Z=3\times 10^{-4}$ yields presented here 
are rather different from those of $Z=10^{-4}$ by \citet{siess10}. While in our case we 
find $-0.5<$[O/Fe]$<0$, in the S10 work it is $-2 <$[O/Fe]$< -1$. 
The reason for this difference is twofold. On one hand, we see by comparing the
temperatures reached at the bottom of the envelope of our models with their counterparts in
the S10 compilation that these latter experience a much stronger HBB, owing to the
smaller metallicity. Also, in analogy with what found in \citet{vd11}, we stress the
importance of the different treatment of mass loss: the Bl\"ocker prescription used here leads to 
higher rates compared to the \citet{VW93} treatment used by \citet{siess10}; our 
models loose their envelope before a great reduction of the surface oxygen is achieved.
This is also clear in the difference in the number of TPs, which exceeds 2000 in S10, while
it barely reaches 50 in our SAGB models.
A confirmation of this comes from the results from computations of models of $6M_{\odot}$ 
and $7M_{\odot}$ where the free parameter entering the Bl\"ocker formula was reduced to 
$\eta_R=0.005$, simulating a reduction of $75\%$ of the mass loss rate: the corresponding 
oxygen yields, indicated with open squares in the right panel of Fig.~\ref{cno}, confirm 
that more oxygen--poor ejecta are produced when the mass loss rate is decreased.

Among the various elements involved in p--capture nucleosynthesis, the evolution of the surface
magnesium, and the amount of this element in the ejecta, is the most sensitive to the
details of the modeling of the AGB phase. This can be seen in the right panel of Fig.~\ref{mgal},
where our yields are compared with those by \citet{karakas10} and \citet{siess10}. 
In analogy with the CNO elements, the yields corresponding to the $Z=8\times 10^{-3}$ 
chemistry are similar, whereas those for the low metallicity models are extremely
different. For $M\leq 3.5M_{\odot}$ an increase in the magnesium content of the ejecta is
found both in our models and in those by \citet{karakas10}; these latter predict a larger 
enrichment in magnesium, partly due to the smaller metallicity adopted.
The magnesium yields of more massive stars are completely different: while our models,
experiencing strong HBB, show a reduction of the initial magnesium, up to --0.6 dex 
at $\sim 5M_{\odot}$, the K10 yields are magnesium--rich,
confirming the difference in the efficiency of the HBB experienced, and in the
relative role played by HBB and TDU. Both our and K10 models are
found to be Al--rich, as a consequence of the conversion of magnesium to aluminium
via p-capture during the interpulse phase.

In the SAGB domain, the differences with respect to the models by \citet{siess10}
reflect the situation already found in the analysis of the oxygen content of the ejecta.
Because of the smaller mass loss experienced, the yields by \citet{siess10} are
more Mg--poor, because there is more time available to destroy the surface magnesium
via HBB.

\section{Conclusions}
We present and discuss new models of stars of intermediate mass, evolved during the
AGB phase, characterized by the occurrence of a series of Thermal Pulses. We also focus on the
SAGB evolution, experienced by models with mass $6M_{\odot} < M < 8M_{\odot}$, that
develop a core of oxygen and neon. These results complete previous explorations from
our group, based on a single metallicity, $Z=10^{-3}$, and extend to metallicities
typical of low--Z Globular Clusters, i.e. $Z=3\times 10^{-4}$, and to a chemistry typical
of substantially higher metallicity clusters, $Z=8\times 10^{-3}$.

In agreement with previous investigations, we find that massive models with $M\geq 3M_{\odot}$
experience HBB, whereas in lower--mass structures the only mechanism active in changing
the surface chemistry is TDU. The ejecta of low--mass AGBs are enriched in the overall 
C+N+O content, and also in magnesium, whereas the more massive models will reverse into 
the interstellar medium gas contaminated essentially by p--capture nucleosynthesis, with 
the depletion of the surface oxygen and magnesium, and the increase in the sodium and 
aluminium content.

The extent of the HBB is found to be strongly sensitive to metallicity:
low--Z, massive AGBs (with $Z=3\times 10^{-4}$, roughly corresponding to [Fe/H]$=-2$)
reach very large temperatures at the bottom of the surface convective mantle, 
exceeding 100MK. The corresponding yields show a small oxygen content, up to $\sim 20$ 
times lower than in the initial mixture, and a magnesium depletion of a factor $\sim 5$. 
Aluminium is increased by a factor $\sim 10-20$, which is the highest abundance 
achievable within the present schematization; this is because at very large temperatures
the production and destruction channels compensate: at these T's, silicon synthesis is 
expected. The oxygen and sodium in the ejecta are correlated, because for $T > 80$MK 
oxygen depletion is accompanied  by the destruction of the sodium previously accumulated 
by the second dredge--up, and further increased in the early AGB phase, by proton--capture 
on $^{22}$Ne nuclei.

The $Z=8\times 10^{-3}$ massive AGB models experience only a modest HBB, thus the ejecta are
expected to produce much less contaminated ejecta: magnesium is hardly touched by the
HBB nucleosynthesis, whereas the depletion of oxygen barely exceeds a factor $\sim 2$.

In the low--mass regime, where the surface chemistry in unaffected by HBB, the results
presented here are in good agreement with other investigations in the literature, although
the extent of the carbon enrichment is sensitive to the details of the treatment of the
convective borders. For more massive models, the results depend on the combined effects of
the description of convection and on the mass loss treatment. While the enhancement of
nitrogen and aluminium appear as rather robust, the extent of magnesium and oxygen depletion
are strongly model dependent, as the sodium content, which is also affected by the
uncertainties in the relevant cross--sections.

Two main results found in the investigations by \citet{vd11} are confirmed here, 
independently of metallicity:

\begin{itemize}

\item{The helium content of the ejecta increases with mass, reaching $Y \sim 0.38$
in the more massive models, in the SAGB regime.}

\item{The models showing the most extreme chemistry are those with mass $M \sim 6M_{\odot}$,
at the edge between the AGB and SAGB regimes. This is due to the large mass loss experienced
by SAGBs, that loose their envelopes before a very advanced HBB nucleosynthesis is
experienced.}

\end{itemize}

\section*{Acknowledgments}
The authors are indebted to the referee, Achim Weiss, for the careful reading of the manuscript,
and for the competent comments and suggestions, that contributed to increase the quality
of this work. This work was partially funded by the PRIN INAF 2009 "Formation and Early 
Evolution of Massive Star Clusters".

\end{document}